\pdfoutput=1
\documentclass[final,3p,times]{elsarticle}
\journal{JLAMP}

\usepackage{float,hyperref}
\usepackage{algorithm,algpseudocode,amsmath,framed,multicol,pifont,tikz}
\usetikzlibrary{arrows.meta,shadows}
\tikzset{
  auto,
  >={Stealth[length=4pt,width=3pt]},
  invclip/.style={
    clip,
    insert path={{[reset cm](-16383.99999pt,-16383.99999pt) rectangle (16383.99999pt,16383.99999pt)}}
  }
}

\algrenewtext{EndWhile}{\algorithmicend}
\algrenewtext{EndIf}{\algorithmicend}
\algrenewcommand{\alglinenumber}[1]{\footnotesize #1}

\newcommand{\astop}{{}^*}
\newcommand{\cp}[1]{\overline{#1}}
\newcommand{\cplop}{\cp{\rule{0em}{1.2ex}~\,}}
\newcommand{\trans}[1]{{{#1}^{\sf T}}}
\newcommand{\comp}{\mathbin{;}}
\newcommand{\I}{\mathsf{I}}
\renewcommand{\O}{\mathsf{O}}
\renewcommand{\L}{\mathsf{L}}
\newcommand{\uni}{\cup}
\newcommand{\isc}{\cap}
\newcommand{\rleq}{\subseteq}
\newcommand{\spt}[1]{\mathsf{start}(#1)}
\newcommand{\ept}[1]{\mathsf{end}(#1)}
\renewcommand{\iff}{\Leftrightarrow}
\renewcommand{\implies}{\Rightarrow}

\newcommand{\pfinite}{\ding{192}}
\newcommand{\pininf}{\ding{193}}
\newcommand{\poutinf}{\ding{194}}
\newcommand{\pdblinf}{\ding{195}}
\newcommand{\pcycle}{\ding{196}}
\newcommand{\pempty}{\ding{197}}

\newcommand{\Input}{\mathbf{input}}
\newcommand{\Output}{\mathbf{output}}
\newcommand{\Point}{\mathsf{choosePoint}}
\newcommand{\Atom}{\mathsf{chooseAtom}}
\newcommand{\Min}{\mathsf{min}}
\newcommand{\Termpath}{\mathsf{termPath}}
\newcommand{\Cycle}{\mathsf{cycle}}
\newcommand{\Ispoint}{\mathsf{point}}
\newcommand{\Isatom}{\mathsf{atom}}
\newcommand{\Pre}{\mathsf{Pre}}
\newcommand{\Post}{\mathsf{Post}}
\newcommand{\Inv}{\mathsf{Inv}}

\newtheorem{lem}{Lemma}[section]
\newtheorem{thm}[lem]{Theorem}
\newtheorem{cor}[lem]{Corollary}
\newtheorem{defi}[lem]{Definition}

\usepackage[normalem]{ulem}
\usepackage{color}

\begin{document}

\title{Relational Characterisations of Paths}

\author[CAU]{Rudolf Berghammer}
\author[KAG]{Hitoshi Furusawa}
\author[CAN]{Walter Guttmann}
\author[DAT,NSW]{Peter H\"ofner}

\address[CAU]{Institut f\"ur Informatik, Christian-Albrechts-Universit\"at zu Kiel, Germany}
\address[KAG]{Department of Mathematics and Computer Science, Kagoshima University, Japan}
\address[CAN]{Department of Computer Science and Software Engineering, University of Canterbury, New Zealand}
\address[DAT]{Data61, CSIRO, Sydney, Australia}
\address[NSW]{Computer Science and Engineering, University of New South Wales, Australia}


\begin{abstract}
  Binary relations are one of the standard ways to encode, characterise and reason about graphs.
  Relation algebras provide equational axioms for a large fragment of the calculus of binary relations.
  Although relations are standard tools in many areas of mathematics and computing, researchers usually fall back to point-wise reasoning when it comes to arguments about paths in a graph.
  We present a purely algebraic way to specify different kinds of paths in relation algebras.
  We study the relationship between paths with a designated root vertex and paths without such a vertex.
  Since we stay in first-order logic this development helps with mechanising proofs.
  To demonstrate the applicability of the algebraic framework we verify the correctness of three basic graph algorithms.
  All results of this paper are formally verified using the interactive proof assistant Isabelle/HOL.
\end{abstract}


\maketitle

\thispagestyle{plain}

\section{Introduction}

Paths are a fundamental concept in graph theory and its applications.
Many textbooks define a path in a directed graph to be a sequence of vertices such that successive vertices in the sequence are connected by an edge of the graph; see \cite{Harary1969,CormenLeisersonRivest1990,Berge2001}.
A simple path is one whose vertices are distinct; a cycle is one whose first and last vertices are identical.
There are variations in terminology, but definitions of paths are mostly based on sequences.

An alternative approach is to define a path as a subgraph of edges that is connected such that every vertex has at most one incoming edge and at most one outgoing edge; this is considered, for example, in \cite{Tinhofer1976,Diestel2005}.
The aim of the present paper is to derive a theory of paths in directed graphs based on this alternative definition and to show how it can be used to verify the correctness of graph algorithms.
Paths according to this definition correspond to simple paths when considered as sequences of vertices; this paper is not concerned with more general kinds of paths.

The main motivation for using subgraphs instead of sequences is that the former allow us to reason about paths as special kinds of graphs using the well-established framework of relation algebra.
Specifically, the edge set $E$ of a directed graph is a subset of the Cartesian product $A \times A$, where $A$ is the set of vertices, and therefore $E$ is a (homogeneous) relation on the set $A$.
Hence both paths and graphs correspond to relations, and relational concepts and methods can be used to work with them in a unified setting.
We are particularly interested in exploiting the algebraic structure of relations, in order to concisely express properties of graphs and reason about graphs using equational, calculational proofs.
Such proofs can be easily mechanised and formally verified using automated theorem provers and interactive proof assistants \cite{HoefnerStruth2008,BerghammerStruth2010,BerghammerEtAl2014,FosterStruthWeber2011,Relation_Algebra-AFP,Pous2016}.

Relation algebras have been proposed by De Morgan, Peirce, Schr\"oder and Tarski to express a rich fragment of first-order logic algebraically without using variables and quantifiers \cite{Tarski1941}; recent textbooks are \cite{HirschHodkinson2002,Maddux2006}.
Besides for logical foundations, relational methods have been used for program analysis \cite{DesharnaisMoellerStruth2011,Guttmann2016a}, refinement \cite{Wright2004}, databases \cite{OkumaKawahara2000}, preference modelling \cite{Schmidt08,MoellerEtAl12}, algorithm development \cite{FriasAguayoNovak1993,BackhouseEtAl94,Berghammer1999,Guttmann2018b} and many other applications; for example, see \cite{COST1,COST2,ScolloFrancoManca06,Schmidt12,Mueller12,Moeller13}.

\pagebreak

Relations and relation algebras have been used for a wide range of topics in graph theory \cite{SchmidtStroehlein1993,BerghammerHoffmann2001,Schmidt2011,BerghammerHoefnerStucke2015,Glueck2017}.
By defining paths as subgraphs, we can reuse the well-developed theory of relation algebras and reason about paths without resorting to variables and quantifiers.

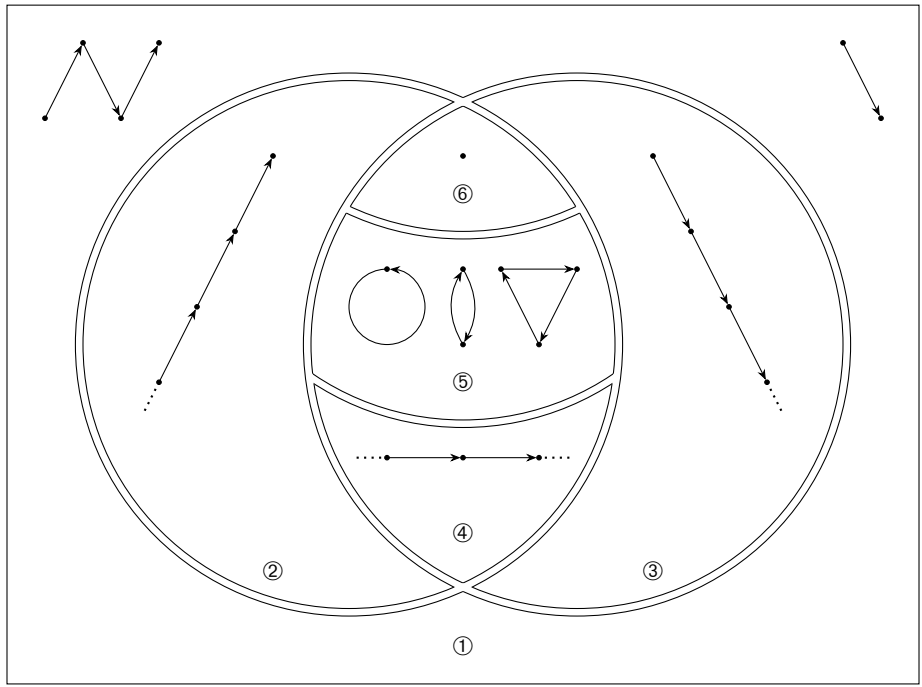
\begin{figure}
  \begin{center}
  \begin{tikzpicture}
    \draw (0,0) rectangle (12,9);

    \begin{pgfinterruptboundingbox} 

    \begin{scope}
      \path[invclip] (7.5,4.5) circle [radius=3.6cm];
      \draw (4.5,4.5) circle [radius=3.6cm];
    \end{scope}
    \begin{scope}
      \path[invclip] (4.5,4.5) circle [radius=3.6cm];
      \draw (7.5,4.5) circle [radius=3.6cm];
    \end{scope}

    \begin{scope}
      \path[invclip] (7.5,4.5) circle [radius=3.6cm];
      \draw (4.5,4.5) circle [radius=3.5cm];
    \end{scope}
    \begin{scope}
      \clip (4.5,4.5) circle [radius=3.5cm];
      \draw (7.5,4.5) circle [radius=3.6cm];
    \end{scope}

    \begin{scope}
      \path[invclip] (4.5,4.5) circle [radius=3.6cm];
      \draw (7.5,4.5) circle [radius=3.5cm];
    \end{scope}
    \begin{scope}
      \clip (7.5,4.5) circle [radius=3.5cm];
      \draw (4.5,4.5) circle [radius=3.6cm];
    \end{scope}

    \begin{scope}
      \clip (4.5,4.5) circle [radius=3.5cm];
      \path[invclip] (6.0,7.0) circle [radius=3.6cm];
      \draw (7.5,4.5) circle [radius=3.5cm];
    \end{scope}
    \begin{scope}
      \clip (7.5,4.5) circle [radius=3.5cm];
      \path[invclip] (6.0,7.0) circle [radius=3.6cm];
      \draw (4.5,4.5) circle [radius=3.5cm];
    \end{scope}

    \begin{scope}
      \clip (4.5,4.5) circle [radius=3.5cm];
      \clip (6.0,7.0) circle [radius=3.5cm];
      \path[invclip] (6.0,9.5) circle [radius=3.6cm];
      \draw (7.5,4.5) circle [radius=3.5cm];
    \end{scope}
    \begin{scope}
      \clip (7.5,4.5) circle [radius=3.5cm];
      \clip (6.0,7.0) circle [radius=3.5cm];
      \path[invclip] (6.0,9.5) circle [radius=3.6cm];
      \draw (4.5,4.5) circle [radius=3.5cm];
    \end{scope}

    \end{pgfinterruptboundingbox}

    \begin{scope}
      \clip (4.5,4.5) circle [radius=3.5cm];
      \clip (6.0,9.5) circle [radius=3.5cm];
      \draw (7.5,4.5) circle [radius=3.5cm];
    \end{scope}
    \begin{scope}
      \clip (7.5,4.5) circle [radius=3.5cm];
      \clip (6.0,9.5) circle [radius=3.5cm];
      \draw (4.5,4.5) circle [radius=3.5cm];
    \end{scope}

    \begin{scope}
      \clip (4.5,4.5) circle [radius=3.5cm];
      \clip (7.5,4.5) circle [radius=3.5cm];
      \draw (6.0,7.0) circle [radius=3.6cm]
            (6.0,7.0) circle [radius=3.5cm]
            (6.0,9.5) circle [radius=3.6cm]
            (6.0,9.5) circle [radius=3.5cm];
    \end{scope}

    \draw[fill] (5.0,3.0) circle (0.3mm)
                (6.0,3.0) circle (0.3mm)
                (7.0,3.0) circle (0.3mm);
    \path[->] (4.6,3.0) edge [-,dotted,thick] (5.0,3.0)
              (5.0,3.0) edge (5.97,3.0)
              (6.0,3.0) edge (6.97,3.0)
              (7.4,3.0) edge [-,dotted,thick] (7.0,3.0);
    \draw[fill] (2.0,4.0) circle (0.3mm)
                (2.5,5.0) circle (0.3mm)
                (3.0,6.0) circle (0.3mm)
                (3.5,7.0) circle (0.3mm);
    \path[->] (2.0,4.0) edge [-,dotted,thick] (1.8,3.6)
              (2.0,4.0) edge (2.485,4.97)
              (2.5,5.0) edge (2.985,5.97)
              (3.0,6.0) edge (3.485,6.97);
    \draw[fill] (10.0,4.0) circle (0.3mm)
                ( 9.5,5.0) circle (0.3mm)
                ( 9.0,6.0) circle (0.3mm)
                ( 8.5,7.0) circle (0.3mm);
    \path[->] (8.5,7.0) edge (8.985,6.03)
              (9.0,6.0) edge (9.485,5.03)
              (9.5,5.0) edge (9.985,4.03)
              (10.0,4.0) edge [-,dotted,thick] (10.2,3.6);
    \draw[fill] (6.5,5.5) circle (0.3mm)
                (7.5,5.5) circle (0.3mm)
                (7.0,4.5) circle (0.3mm);
    \path[->] (6.5,5.5) edge (7.47,5.5)
              (7.5,5.5) edge (7.015,4.53)
              (7.0,4.5) edge (6.515,5.47);
    \draw[fill] (6.0,5.5) circle (0.3mm)
                (6.0,4.5) circle (0.3mm);
    \path[->] (6.0,5.5) edge [bend left] (6.015,4.53)
              (6.0,4.5) edge [bend left] (5.985,5.47);
    \draw[fill] (5.0,5.5) circle (0.3mm);
    \draw[->] (5.0,5.5) arc [start angle=90,delta angle=356,radius=0.5];
    \draw[fill] (6.0,7.0) circle (0.3mm);
    \draw[fill] (0.5,7.5) circle (0.3mm)
                (1.0,8.5) circle (0.3mm)
                (1.5,7.5) circle (0.3mm)
                (2.0,8.5) circle (0.3mm);
    \path[->] (0.5,7.5) edge (0.985,8.47)
              (1.0,8.5) edge (1.485,7.53)
              (1.5,7.5) edge (1.985,8.47);
    \draw[fill] (11.0,8.5) circle (0.3mm)
                (11.5,7.5) circle (0.3mm);
    \path[->] (11.0,8.5) edge (11.485,7.53);
    \node at (6.0,0.5) {\pfinite};
    \node at (3.5,1.5) {\pininf};
    \node at (8.5,1.5) {\poutinf};
    \node at (6.0,2.0) {\pdblinf};
    \node at (6.0,4.0) {\pcycle};
    \node at (6.0,6.5) {\pempty};
  \end{tikzpicture}
  \end{center}
  \caption{Six disjoint classes of paths}
  \label{fig:class}
\end{figure}

The present paper studies paths using relation-algebraic means.
Its contributions are:
\begin{itemize}
\item[--] Equational characterisations of various classes of paths including cycles, finite paths, one-sided and two-sided infinite paths in Sections \ref{section.paths}--\ref{section.cycles}.
      An overview of the classification is shown in Figure \ref{fig:class} and discussed below.
\item[--] Relation-algebraic specifications and correctness proofs of a number of basic graph algorithms that rely on paths in Section \ref{section.verification}.
\item[--] Relation-algebraic characterisations of paths with designated roots and their equivalence to paths without roots in Section \ref{section.roots}.
\end{itemize}
All concepts, theorems and algorithms described in this paper have been implemented in Isabelle/HOL \cite{NipkowPaulsonWenzel2002}.
All results have been formally verified in this system making heavy use of its integrated automated theorem provers and SMT solvers \cite{PaulsonBlanchette2010,BlanchetteBoehmePaulson2011}.
We omit the proofs, which can be found in the theory files under preparation for the Archive of Formal Proofs and currently available at \url{http://www.csse.canterbury.ac.nz/walter.guttmann/algebra/}.

\medskip
\noindent
Figure \ref{fig:class} shows the classes of paths discussed in this paper:
\begin{itemize}
\item[\pfinite] Finite paths, which have both a start vertex and an end vertex, and finitely many edges.
\item[\pininf] One-sided infinite paths that have an end vertex but no start vertex.
\item[\poutinf] One-sided infinite paths that have a start vertex but no end vertex.
\item[\pdblinf] Two-sided infinite paths, which have neither a start vertex nor an end vertex, but infinitely many edges.
\item[\pcycle] Cycles, which have neither a start vertex nor an end vertex, and at least one but finitely many edges.
\item[\pempty] The empty path, which has neither a start vertex nor an end vertex, and no edges.
\end{itemize}
We refer to unions of these disjoint classes by listing the associated labels; for example, [\pfinite\poutinf] are the paths that have a start vertex.
The left circle of Figure \ref{fig:class} contains the paths without a start vertex [\pininf\pdblinf\pcycle\pempty] and the right circle those without an end vertex [\poutinf\pdblinf\pcycle\pempty].

\section{Relation Algebras}
\label{section.relation-algebra}

In this section we give basic definitions, operations and properties of relations.
Their structure is captured abstractly by relation algebras, which let us characterise important properties of relations in a compact way using equations and inequalities.
We explain how to represent edges, vertices and sets of vertices of a graph as relations.
To describe reachability in graphs, we recall first-order axioms for the Kleene star operation.
Finally, we discuss tool support for mechanising the presented concepts.

\subsection{Relations}
\label{subsection.relations}

A (concrete) binary relation $R$ on a set $A$ is a subset of the Cartesian product $A \times A$, that is, a set of ordered pairs of elements of $A$.
Using the language of graph theory, $A$ is the set of vertices of the graph and $R$ is the set of directed edges.
In this context, we also call $R$ a graph.
Binary relations can furthermore be understood as Boolean matrices with rows and columns indexed by $A$, which corresponds to the adjacency matrix representation of graphs.

Since relations are sets, the union $R \uni S$, intersection $R \isc S$ and complement $\cp{R}$ of relations $R$ and $S$ can be taken such that the set of all binary relations on $A$ forms a Boolean algebra.
The \emph{empty relation} $\O$ on $A$ is the empty set and the \emph{universal relation} $\L$ on $A$ is the full Cartesian product $A \times A$.
The \emph{composition} $R \comp S$ of two relations $R$ and $S$ is the set of all pairs $(a,c) \in A \times A$ such that $(a,b) \in R$ and $(b,c) \in S$ for some $b \in A$.
The \emph{converse} or \emph{transpose} $\trans{R}$ of a relation $R$ is the set of all pairs $(a,b)$ with $(b,a) \in R$.
The \emph{identity relation} $\I$ on $A$ is the set of all pairs $(a,a)$ with $a \in A$.

The structure $(2^{A \times A},\uni,\comp,\cplop,\trans{},\I)$ is called the concrete relation algebra of all binary relations over $A$.
The operations $\isc$, $\O$ and $\L$ can be defined in terms of the other operations.
To discuss the algebraic structure of relations in a more abstract way, binary relations are replaced by arbitrary elements of a carrier set $B$, operations on $B$ are introduced and they are axiomatised by equations as follows.

An \emph{(abstract) relation algebra} is a structure $(B,\uni,\comp,\cplop,\trans{},\I)$ satisfying the axioms \cite{Tarski1941,Maddux2006}
\begin{align*}
    (R \uni S) \uni T & = R \uni (S \uni T)
  & R \uni S & = S \uni R
  & R & = \cp{\cp{R} \uni \cp{S}} \uni \cp{\cp{R} \uni S}
  \\
    (R \comp S) \comp T & = R \comp (S \comp T)
  & (R \uni S) \comp T & = R \comp T \uni S \comp T
  & R \comp \I & = R
  \\
    \trans{\trans{R}} & = R
  & \trans{(R \uni S)} & = \trans{R} \uni \trans{S}
  & \trans{(R \comp S)} & = \trans{S} \comp \trans{R}
  \\
    &
  & \trans{R} \comp \cp{R \comp S} \uni \cp{S} & = \cp{S}
\end{align*}
The first line contains Huntington's axioms for Boolean algebras \cite{Huntington1933a,Huntington1933b}.
The join operation is denoted by $\uni$, based on which the meet can be defined as $R \isc S = \cp{\cp{R} \uni \cp{S}}$.
Since every relation algebra is a Boolean algebra, the set $B$ is partially ordered by $R \rleq S \Leftrightarrow R \uni S = S$ with greatest element $\L = R \uni \cp{R}$ and least element $\O = R \isc \cp{R}$.
In the graph model, $\O$ is the empty graph and $\L$ is the complete graph.
The axioms in the second line give properties of composition $\comp$.
It follows that every relation algebra is a semiring with the two operations $\uni$ and $\comp$.
The axioms in the last two lines specify the operation of converse.
We assume that composition has higher precedence than join and meet and that complement and converse have higher precedence than composition.

It follows that the operations $\uni$, $\isc$, $\comp$ and $\trans{}$ preserve the order $\rleq$ and the operation $\cplop$ reverses the order $\rleq$.
As further examples, we discuss two properties which follow from the above axioms of relation algebras:
\begin{align}
  \label{eq:triple}
  R & \rleq R \comp \trans{R} \comp R \\
  \label{eq:loop_backward_forward}
  R \isc \I & = \trans{R} \isc \I
\end{align}
We interpret these properties in the graph model.
Inequality (\ref{eq:triple}) holds because any edge $(a,b) \in R$ in the graph can be traversed backwards in $\trans{R}$, since $(b,a) \in \trans{R}$, so the edge $(a,b)$ is also contained in the composition $R \comp \trans{R} \comp R$ by going forward, backward and forward again from $a$ via $b$ and $a$ to $b$.
Equality (\ref{eq:loop_backward_forward}) contains an intersection with the identity relation on both sides; it therefore considers only loops in the graph, that is, edges, from a vertex to itself.
It states that such edges are not changed if a graph is transposed, that is, if all its edges are reversed.

Many further properties of relation algebras can be found in textbooks such as \cite{SchmidtStroehlein1993,Maddux2006,Schmidt2011}.

\subsection{Relational Properties}

Compact equational characterisations of special classes of relations can be given in relation algebras.
We use the following properties of functions and orders.

A relation $R$ is \emph{univalent} if $\trans{R} \comp R \rleq \I$ and \emph{total} if $\I \rleq R \comp \trans{R}$.
A relation $R$ is \emph{injective} if $\trans{R}$ is univalent, \emph{surjective} if $\trans{R}$ is total and \emph{bijective} if $R$ is injective and surjective.
A relation $R$ is \emph{irreflexive} if $R \rleq \cp{\I}$ and \emph{symmetric} if $R = \trans{R}$.

For concrete relations, the equivalence of these relation-algebraic specifications and the common logical specifications can be easily derived.
For example, $R \rleq A \times A$ is univalent if and only if for each $a \in A$ there is at most one $b \in A$ such that $(a,b) \in R$.
If $R$ is interpreted as the edge set of a directed graph with vertex set $A$, this means that the out-degree of every vertex is at most $1$.
Similarly, $R$ is total if and only if the out-degree of every vertex is at least $1$.
Conversely, injective and surjective state the same requirements for the in-degree of vertices instead of their out-degree.
Irreflexivity specifies that a graph contains no loops.
Symmetric relations are sometimes used to represent undirected graphs by containing both $(a,b)$ and $(b,a)$ if there is an edge between $a$ and $b$.

A relation $R$ is surjective if and only if $\L \comp R = \L$.
As another consequence of the above properties, we present the following result about injective and surjective relations.
For this, and following results, we assume that variables range over a given relation algebra.

\begin{lem}
  \label{lemma.inj_sur_semi_swap,bij_swap}
  If $P$ is surjective and $R$ is injective, $P \rleq Q \comp R$ implies $R \rleq \trans{Q} \comp P$.
  These two inequalities are equivalent if $P$ and $R$ are bijective.
\end{lem}

\subsection{Vectors, Points and Atoms}
\label{subsection.points}

We now discuss three particular properties that are useful to represent sets of elements as relations.
A relation $v$ is a \emph{vector} if $v = v \comp \L$.
A \emph{point} is a bijective vector.
A relation $x$ is an \emph{atom} if both $x \comp \L$ and $\trans{x} \comp \L$ are points; see \cite{SchmidtStroehlein1993}.

In the matrix model, a vector corresponds to a row-constant matrix.
That is, $v \rleq A \times A$ is a vector if and only if for every $a \in A$, the pair $(a,b)$ is in $v$ either for all $b \in A$ or for none.
Such a relation is used to model the subset of elements of $A$ that are related by $v$ to all elements of $A$.
In the graph model, this can be used to represent sets of vertices.

A point is a vector that is additionally injective and surjective.
In the graph model, this means that the in-degree of every vertex is exactly $1$, so the adjacency matrix contains exactly one row with $1$-entries.
This means that a point represents a set that contains exactly one element.
Such singleton sets obviously correspond to elements of $A$ and can therefore be used to represent individual vertices of graphs.

An atom is a relation consisting of a single pair.
Specifically, if $R$ is an atom and $(a,b) \in R$, then $R \comp \L$ is the point representing the element $a$, and similarly $\trans{R} \comp \L$ represents the element $b$.
Hence an atom corresponds to a single edge in a graph.

The composition $R \comp \L$ is a vector for any relation $R$.
More generally, the composition $R \comp v$ of a relation $R$ and a vector $v$ is again a vector.
It represents the set of vertices from which there are transitions under $R$ into the set represented by $v$.
In the graph model this amounts to the predecessors of the vertices in the set represented by $v$.
Similarly, $\trans{R} \comp v$ is the set of successors of the vertices in the set represented by $v$.

The following result gives an example of a property of vectors.

\begin{lem}
  \label{lemma.vector_meet_comp}
  For vectors $v$, $w$, we have $v \comp \trans{w} = v \isc \trans{w}$.
  In particular $R \comp \L \comp S = R \comp \L \isc \L \comp S$ for all relations $R$, $S$.
\end{lem}

We will use two further properties of relations, which do not follow from the axioms of relation algebras given in Section \ref{subsection.relations}.
The first one is the \emph{Tarski rule}:
\[
  R \neq \O \iff \L \comp R \comp \L = \L
\]
It can be interpreted as follows.
If $R \rleq A \times A$ is not empty, it contains a pair $(a,b)$ for some $a, b \in A$.
Then the vector $R \comp \L$ represents a set that contains $a$; in particular, $R \comp \L$ contains a point.
Hence $\L \comp R \comp \L$ is the universal relation since points are surjective and composition preserves the order $\rleq$.
The backward implication of the Tarski rule is equivalent to $\O \neq \L$, which holds if and only if the set of vertices of a graph is not empty.

Using the Tarski rule it can be shown that a relation $p$ is a point if and only if $p$ is a vector, $p$ is injective and $p \neq \O$.

The second property of relations we use is the \emph{point axiom}.
It states that for each $R \neq \O$ there are points $p$ and $q$ such that $p \comp \trans{q} \rleq R$.
Again, by the assumption there are $a, b \in A$ such that $(a,b) \in R$.
Then points $p$ and $q$ can be chosen so as to represent the vertices $a$ and $b$, respectively.
By Lemma \ref{lemma.vector_meet_comp} we have $p \comp \trans{q} = p \isc \trans{q}$, whence this expression intersects row $a$ and column $b$ of the matrix to yield the entry $(a,b)$.
In particular, it follows that $p \comp \trans{q}$ is an atom.
To summarise, the point axiom states that every non-empty graph contains an edge, which is obvious in the graph model but an independent property in relation algebras.

\subsection{Reflexive Transitive Closure}

To describe reachability in graphs we expand relation algebras with an operation for the reflexive transitive closure.
A \emph{Kleene relation algebra} is a structure $(M,\uni,\comp,\cplop,\trans{},\astop,\I)$ such that $(M,\uni,\comp,\cplop,\trans{},\I)$ is a relation algebra and the Kleene star operation $\astop$ satisfies the following axioms:
\begin{align*}
  \I \uni R \comp R^* & \rleq R^* & S \uni R \comp Q \rleq Q & \implies R^* \comp S \rleq Q \\
  \I \uni R^* \comp R & \rleq R^* & S \uni Q \comp R \rleq Q & \implies S \comp R^* \rleq Q
\end{align*}
Relation algebras with transitive closure have been studied, for example, in \cite{Ng1984}; Kleene algebras have been studied in \cite{Conway1971}; the axioms above were proposed in \cite{Kozen1994}.

The operation $\astop$ models reflexive transitive closure, from which the transitive closure is easily obtained by $R^+ = R \comp R^*$.
Hence an edge $(a,b)$ is in $R^*$ if and only if there is a path from vertex $a$ to vertex $b$ in the graph $R$, and a similar statement holds for $R^+$ and non-empty paths.
A relation $R$ is \emph{acyclic} if $R^+$ is irreflexive, that is, there is no non-empty path from a vertex to itself.

For example, we obtain the following consequences:
\begin{align*}
  R^* \uni S^* & = R^+ \uni S^* \\
  \trans{R}^* & = \trans{R^*}
\end{align*}

In the remainder of this paper we work in Kleene relation algebras that satisfy the Tarski rule and the point axiom.
We note, however, that many of the following results do not require the Tarski rule and/or the point axiom but hold in more general structures, as can be seen in our Isabelle/HOL implementation.

\subsection{Formalisation and Mechanisation}

As shown above, relation algebras facilitate compact equational specifications of relational properties and support first-order axioms for reachability, which is a key notion for dealing with paths.
All relation-algebraic statements can be expressed in logic, but doing so would introduce complex formulas with nested quantifiers.
For example, a relation $R \subseteq A \times A$ is surjective if and only if
$
  \forall y \in A : \exists x \in A : (x,y) \in R
$
and $R$ is injective if and only if
$
  \forall y \in A : \forall x \in A : \forall w \in A : (x,y) \in R \wedge (w,y) \in R \implies x = w
$
In relation algebras, these properties are expressed as $\I \rleq \trans{R} \comp R$ and $R \comp \trans{R} \rleq \I$, respectively.
Like here, relation algebras often yield simpler, more modular and more concise specifications and proofs.
This is one reason why we formalise our work using relation algebras rather than logic.

Due to their compact quantifier-free form, statements given in relation algebras can be more easily tackled by automated and interactive theorem provers.
Off-the-shelf automated theorem provers such as E \cite{Schulz2013} and Prover9 \cite{Prover9} are performing well when proving statements containing relational expressions \cite{HoefnerStruth2008,DangHoefner2008}.
There are also special-purpose first-order proof systems for relation algebras; for example, see \cite{Maddux1983,MacCaullOrlowska2002}.
These systems, however, usually fail when proving complex properties of relation algebras since the search space becomes too large.

To overcome this deficiency, interactive proof assistants such as Coq \cite{Bertot2004} and Isabelle/HOL \cite{NipkowPaulsonWenzel2002} can be used.
Both systems have been successfully used in the context of relation algebras \cite{Pous2016,Relation_Algebra-AFP}.
Key differences are that Coq has a more expressive type system whereas Isabelle/HOL has better automation support.
In particular, the Sledgehammer tool \cite{Meng2008,PaulsonBlanchette2010} integrates first-order automatic theorem provers and SMT solvers to discharge goals arising in interactive Isabelle/HOL proofs.
Since our work does not require an elaborate type system but greatly benefits from automation we have implemented all results of this paper in Isabelle/HOL.
Proofs are omitted and can be found in the accompanying Isabelle theories.
To improve readability, most of our proofs are written in Isabelle/Isar (Intelligible Semi-Automated Reasoning) \cite{Wenzel2002}.

\section{Paths}
\label{section.paths}

In this section we start our endeavour to characterise classes of paths.
One of the preconditions of a path between two points is that the points are connected.

\begin{defi}
  \label{definition.connected}
  A relation $R$ is \emph{connected} if $R \comp \L \comp R \rleq R^* \uni \trans{R}^*$.
\end{defi}

The expression $R \comp \L \comp R$ is equivalent to $R \comp \L \isc \L \comp R$ by Lemma \ref{lemma.vector_meet_comp} and describes all pairs $(a,b)$ such that vertex $a$ has an outgoing edge in $R$ and vertex $b$ has an incoming edge in $R$.
The inequality requires that there must be a path from $a$ to $b$ in $R$ or a path from $b$ to $a$ in $R$.
Hence the definition states that any vertex with a successor in $R$ must be connected by a path to any vertex with a predecessor in $R$ either by going forwards ($R^*$) or backwards ($\trans{R}^*$).

Connected relations do not characterise paths since they allow, for instance, the following graphs:
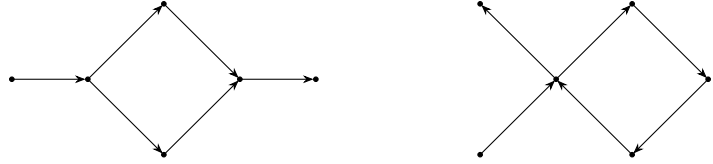
\begin{center}
  \begin{tikzpicture}
    \draw[fill] (0.0,1.0) circle (0.3mm)
                (1.0,1.0) circle (0.3mm)
                (2.0,2.0) circle (0.3mm)
                (2.0,0.0) circle (0.3mm)
                (3.0,1.0) circle (0.3mm)
                (4.0,1.0) circle (0.3mm);
    \path[->] (0.0,1.0) edge (0.97,1.0)
              (1.0,1.0) edge (1.985,1.985)
              (1.0,1.0) edge (1.985,0.015)
              (2.0,2.0) edge (2.985,1.015)
              (2.0,0.0) edge (2.985,0.985)
              (3.0,1.0) edge (3.97,1.0);
  \end{tikzpicture}
  \hskip20mm
  \begin{tikzpicture}
    \draw[fill] (0.0,0.0) circle (0.3mm)
                (0.0,2.0) circle (0.3mm)
                (1.0,1.0) circle (0.3mm)
                (2.0,0.0) circle (0.3mm)
                (2.0,2.0) circle (0.3mm)
                (3.0,1.0) circle (0.3mm);
    \path[->] (0.0,0.0) edge (0.985,0.985)
              (1.0,1.0) edge (1.985,1.985)
              (2.0,2.0) edge (2.985,1.015)
              (3.0,1.0) edge (2.015,0.015)
              (2.0,0.0) edge (1.015,0.985)
              (1.0,1.0) edge (0.015,1.985);
  \end{tikzpicture}
\end{center}
We additionally have to prevent paths from forking into several successors and from merging several predecessors.

\begin{defi}
  A relation $R$ is a \emph{path} if $R$ is injective, univalent and connected.
\end{defi}

Paths in this sense encompass all classes shown in Figure \ref{fig:class} [\pfinite\pininf\poutinf\pdblinf\pcycle\pempty], and nothing else.
In particular, a path can be finite, one-sided infinite, two-sided infinite, a cycle or even the empty relation.
In the remainder of the paper we will look at each class separately.

\begin{thm}
  \label{theorem.conv_path}
  $R$ is connected if and only if $\trans{R}$ is connected.
  $R$ is a path if and only if $\trans{R}$ is a path.
\end{thm}

The following theorem shows equivalent formalisations of connectivity.
Statement (\ref{theorem.connectivity.1}) is the property given in Definition \ref{definition.connected}.

\begin{thm}
  For a univalent and injective relation $R$, the following properties are equivalent:
  \begin{multicols}{2}
  \begin{enumerate}
  \item $R \comp \L \comp R \rleq R^* \uni \trans{R}^*$
        \label{theorem.connectivity.1}
  \item $R \comp \L \comp \trans{R} \rleq R^* \uni \trans{R}^*$
  \item $\trans{R} \comp \L \comp R \rleq R^* \uni \trans{R}^*$
  \item $\trans{R} \comp \L \comp \trans{R} \rleq R^* \uni \trans{R}^*$
  \columnbreak
  \item $R^+\comp \L \comp R^+ \rleq R^* \uni \trans{R}^*$
  \item $R^+\comp \L \comp \trans{R}^+ \rleq R^* \uni \trans{R}^*$
  \item $\trans{R}^+ \comp \L \comp R^+ \rleq R^* \uni \trans{R}^*$
  \item $\trans{R}^+ \comp \L \comp \trans{R}^+ \rleq R^* \uni \trans{R}^*$
  \end{enumerate}
  \end{multicols}
\end{thm}

\section{Start Points and End Points}
\label{section.start-end-points}

Many paths have special vertices, namely a start point (root) and an end point (sink).
A start point is a vertex without predecessor, and an end point is a vertex without successor.
This is captured by the following definition.

\begin{defi}
  The \emph{start points} of a relation $R$ are given by $\spt{R} = R \comp \L \isc \cp{\trans{R} \comp \L}$ and its \emph{end points} are given by $\ept{R} = \trans{R} \comp \L \isc \cp{R \comp \L}$.
\end{defi}

As above, $R \comp \L$ describes all vertices having a successor, and $\trans{R} \comp \L$ those vertices with a predecessor.
Hence, $R \comp \L \isc \cp{\trans{R} \comp \L}$ characterises the vertices with at least one successor and no predecessors, which are just the start points of a path.

Immediate consequences are $\spt{R} = \ept{\trans{R}}$ and $R \comp \spt{R} = \trans{R} \comp \ept{R} = \O$.
Of course, not every path has such distinguished elements.
In Figure \ref{fig:class}, the classes [\pfinite\poutinf] comprise the paths with start points.
Similarly, classes [\pfinite\pininf] are the paths with end points.
A cycle or a two-sided infinite path has neither start points nor end points, in which case $\spt{R} = \ept{R} = \O$.
The following result shows that every path has at most one start point and at most one end point.

\begin{thm}
  \label{theorem.start_point_at_most_one}
  Let $R$ be a path.
  Then $\spt{R}$ and $\ept{R}$ are injective.
\end{thm}

It follows that start points and end points are almost points in the relation-algebraic sense.
The only exception is the empty relation as the following result states.

\begin{cor}
  Let $R$ be a path.
  Then $\spt{R} = \O$ or $\spt{R}$ is a point and, similarly, $\ept{R} = \O$ or $\ept{R}$ is a point.
\end{cor}

The following consequences characterise the existence of start and end points by (in)equalities.
We also give the corresponding classes in Figure \ref{fig:class}.

\begin{cor}
  \label{corollary.start-end-point-iff}
  Let $R$ be a path.
  Then
  \begin{enumerate}
  \item $\spt{R} \neq \O$ if and only if $\L = \cp{\L \comp R} \comp R \comp \L$ {\upshape[\pfinite\poutinf]}.
  \item $\spt{R} = \O$ if and only if $R \comp \L \rleq \trans{R} \comp \L$ {\upshape[\pininf\pdblinf\pcycle\pempty]}.
  \item $\ept{R} \neq \O$ if and only if $\L = \L \comp R \comp \cp{R \comp \L}$ {\upshape[\pfinite\pininf]}.
  \item $\ept{R} = \O$ if and only if $\trans{R} \comp \L \rleq R \comp \L$ {\upshape[\poutinf\pdblinf\pcycle\pempty]}.
  \item $\spt{R} \neq \O$ and $\ept{R} \neq \O$ if and only if $\L = \cp{\L \comp R} \comp R \comp \L \isc \L \comp R \comp \cp{R \comp \L}$ {\upshape[\pfinite]}.
  \item $\spt{R} = \ept{R} = \O$ if and only if $R \comp \L = \trans{R} \comp \L$ {\upshape[\pdblinf\pcycle\pempty]}.
  \end{enumerate}
\end{cor}

It follows that a path has a start point if and only if the converse path has an end point.

The simplest kind of path that has both a start point and an end point is a single edge.
It can be constructed from two (relational) points as shown by the following result.

\begin{lem}
  Let $p$ and $q$ be points.
  Then $p \comp \trans{q}$ is a path.
  If $p \neq q$, then $\spt{p \comp \trans{q}} = p$ and $\ept{p \comp \trans{q}} = q$.
\end{lem}

The start point of the constructed edge is the vertex represented by the point $p$; its end point is the one represented by $q$.
If $p$ and $q$ coincide, $\spt{p \comp \trans{q}} = \ept{p \comp \trans{q}} = \O$ since the result is a loop.

A path with no start point and no end point [\pdblinf\pcycle\pempty] need not be infinite; it can also be the empty relation or a cycle.
We will elaborate on this distinction in Section \ref{section.cycles}.

Sometimes we wish to include the empty relation in our reasoning.
If a path has either a start point or is empty, we call it \emph{backward terminating} [\pfinite\poutinf\pempty].
Symmetrically, we call a path \emph{forward terminating} if it has end points or is empty [\pfinite\pininf\pempty].
If a path is both forward terminating and backward terminating, we call it \emph{terminating} [\pfinite\pempty].
The following result shows how to express each of these properties as an inequality.

\begin{thm}
  Let $R$ be a path.
  Then the following properties are equivalent and each characterises backward termination.
  \begin{enumerate}
  \item $\L = \cp{\L \comp R} \comp R \comp \L$ or $R = \O$
  \item $R \rleq \cp{\L \comp R} \comp R \comp \L$
  \item $R \rleq \cp{\L \comp R} \comp R^*$
  \item $R \rleq \trans{R}^* \comp \cp{\trans{R} \comp \L}$
  \end{enumerate}
  Moreover, each of the following properties is equivalent to forward termination.
  \begin{enumerate}
  \setcounter{enumi}{4}
  \item $\L = \L \comp R \comp \cp{R \comp \L}$ or $R = \O$
  \item $R \rleq \L \comp R \comp \cp{R \comp \L}$
  \item $R \rleq R^* \comp \cp{R \comp \L}$
  \item $R \rleq \cp{\L \comp \trans{R}} \comp \trans{R}^*$
  \end{enumerate}
  Finally, each of the following properties is equivalent to termination.
  \begin{enumerate}
  \setcounter{enumi}{8}
  \item $\L = \cp{\L \comp R} \comp R \comp \L \isc \L \comp R \comp \cp{R \comp \L}$ or $R = \O$
  \item $R \rleq \cp{\L \comp R} \comp R \comp \L \isc \L \comp R \comp \cp{R \comp \L}$
  \item $R \rleq \cp{\L \comp R} \comp R^* \isc R^* \comp \cp{R \comp \L}$
  \item $R \rleq \trans{R}^* \comp \cp{\trans{R} \comp \L} \isc \cp{\L \comp \trans{R}} \comp \trans{R}^*$
  \end{enumerate}
\end{thm}

It follows that a path is backward terminating if and only if its converse is forward terminating.
Moreover a path is terminating if and only if its converse is terminating.

If the end point of a path $R$ and the start point of a path $S$ coincide, the paths can be concatenated.
It has to be guaranteed, however, that the two paths do not cross each other.
The condition used in the following result allows that the start point of $R$ and the end point of $S$ coincide, if they exist, but requires all other vertices of $R$ and $S$ to be distinct.

\begin{thm}
  \label{thm:path_concat}
  Let $R$ be a forward terminating path and let $S$ be a backward terminating path with $\ept{R} = \spt{S}$ and $R \comp \L \isc (\trans{R} \comp \L \uni S \comp \L) \isc \trans{S} \comp \L = \O$.
  Then $R \uni S$ is a path.
  Moreover, $\spt{R \uni S} \rleq \spt{R}$ and $\ept{R \uni S} \rleq \ept{S}$.
\end{thm}

If additionally $\spt{R} = \ept{S}$ are non-empty and hence points, path concatenation creates a cycle.
As a consequence we only obtain inequalities in the second part of the previous theorem.
By strengthening its assumption to $R \comp \L \isc \trans{S} \comp \L = \O$ we can exclude the creation of a cycle.
This means that $R$ and $S$ only meet at the end point of $R$ and the start point of $S$.

\begin{lem}
  The strengthened assumption $R \comp \L \isc \trans{S} \comp \L = \O$ is equivalent to the conjunction of the assumption $R \comp \L \isc (\trans{R} \comp \L \uni S \comp \L) \isc \trans{S} \comp \L = \O$ of Theorem \ref{thm:path_concat} and $\spt{R} \isc \ept{S} = \O$.
\end{lem}

Under this assumption, it is possible to determine the start and end points of the composed path $R \uni S$ according to the following result.

\begin{thm}
  Let $R$ be a forward terminating path and let $S$ be a backward terminating path such that $\ept{R} = \spt{S}$ and $R \comp \L \isc \trans{S} \comp \L = \O$.
  Then $\spt{R \uni S} = \spt{R}$ and $\ept{R \uni S} = \ept{S}$.
\end{thm}

Rather than looking at path concatenation, one can also consider path restriction.
Given a path $R$ and an arbitrary point $p$, we construct a path with start point $p$ by following the edges of $R$.

\begin{thm}
  Let $p$ be a point and let $R$ be a path.
  Then $\trans{R}^* \comp p \isc R$ is a path.
  Moreover, $\spt{\trans{R}^* \comp p \isc R} \rleq p$, and $\ept{\trans{R}^* \comp p \isc R} \rleq \ept{R}$.
\end{thm}

The second statement shows that either $p$ is the start point of the path, or there is no start point at all, which means that the new path is a cycle or empty.
The expression $\trans{R}^* \comp p$ represents the set of predecessors of the point $p$ under the relation $\trans{R}^*$, which are the successors of $p$ under $R^*$, which is the set of vertices reachable from $p$ by a path in $R$.
Intersecting with $R$ keeps the edges of $R$ that start in such a vertex.

The additional assumption $p \rleq R \comp \L$ assures that the point lies on the path $R$ and has at least one successor.
In that case $\spt{\trans{R}^* \comp p \isc R} = p$, and $\ept{\trans{R}^* \comp p \isc R} = \ept{R}$.

\section{Cycles}
\label{section.cycles}

By Corollary \ref{corollary.start-end-point-iff} a path $R$ has no start point and no end point if and only if $R \comp \L = \trans{R} \comp \L$.
In this section we distinguish three kinds of paths satisfying this property: the empty path, cycles and two-sided infinite paths.
A cycle is obtained by assuming a strong connectivity requirement, given in the following definition.

\begin{defi}
  A path $R$ is a \emph{cycle} if $R^* = \trans{R}^*$.
\end{defi}

Cycles in this sense encompass classes [\pcycle\pempty] of Figure \ref{fig:class}.
An immediate consequence is that a relation is a cycle if and only if its converse is a cycle.

The additional property $R^* = \trans{R}^*$ specifies that if a vertex $b$ is reachable from a vertex $a$ in $R$, also $a$ is reachable from $b$ in $R$.
In this case, $a$ and $b$ belong to the same strongly connected component of $R$.
The following result gives equivalent ways to express this property and some consequences.
For example, $R \rleq \trans{R}^+$ states that for every edge $(a,b) \in R$ vertex $b$ is reachable from vertex $a$ by a non-empty path backwards in $R$.

\begin{thm}
  \label{theorem.many_strongly_connected}
  The following properties are equivalent:
  \begin{multicols}{4}
  \begin{enumerate}
  \item $R^* = \trans{R}^*$
  \item $R^+ = \trans{R}^+$
  \item $\trans{R} \rleq R^*$
  \item $\trans{R} \rleq R^+$
  \item $R \rleq \trans{R}^*$
  \item $R \rleq \trans{R}^+$
  \item $R^* \comp \trans{R} \rleq R^+$
  \item $\trans{R} \comp R^* \rleq R^+$
  \end{enumerate}
  \end{multicols}
  \noindent
  Each of them implies the following properties:
  \begin{multicols}{3}
  \begin{enumerate}
  \setcounter{enumi}{8}
  \item $R \comp \L = \trans{R} \comp \L$
  \item $R \comp \trans{R} \rleq R^+$
  \item $\trans{R} \comp R \rleq R^+$
  \item $R \comp \trans{R} \comp R^* \rleq R^+$
  \item $R^* \comp \trans{R} \comp R \rleq R^+$
  \vfill\null\columnbreak
  \item $R^* \comp R \comp \trans{R} \rleq R^+$
  \item $\trans{R} \comp R \comp R^* \rleq R^+$
  \vfill\null
  \end{enumerate}
  \end{multicols}
  \noindent
  If $R$ is univalent, each of (10), (12) and (14) is equivalent to property (1).
  If $R$ is injective, each of (11), (13) and (15) is equivalent to property (1).
\end{thm}

Theorem \ref{theorem.many_strongly_connected} remains valid if any of the inequalities in (7), (8), (12)--(15) are replaced with equalities.

As witnessed by the identity relation $\I$, the property $R^* = \trans{R}^*$ alone does not imply that the relation $R$ contains just one strongly connected component.
This results from the combination with the connectivity requirement given in Definition \ref{definition.connected}.
The following result expresses the combination of these two properties as a single (in)equality.

\begin{thm}
  \label{theorem.one_strongly_connected}
  The following properties are equivalent:
  \begin{multicols}{2}
  \begin{enumerate}
  \item $R$ is connected and $R^* = \trans{R}^*$
  \item $\trans{R} \comp \L \comp \trans{R} \rleq R^*$
  \item $\trans{R} \comp \L \comp \trans{R} \rleq R^+$
  \item $\trans{R} \comp \L \comp \trans{R} = R^+$
  \item $\trans{R} \comp \L \comp R = R^+$
  \item $R \comp \L \comp \trans{R} = R^+$
        \label{schmidt-simple-circuit}
  \end{enumerate}
  \end{multicols}
  \noindent
  They imply each of the following properties:
  \begin{enumerate}
  \setcounter{enumi}{6}
  \item $\trans{R} \comp \L \comp R \rleq R^+$
  \item $R \comp \L \comp \trans{R} \rleq R^+$
  \item $R \comp \L \comp R \rleq R \comp R^+$
  \item $R \comp \L \comp R \rleq R^+$
  \end{enumerate}
  If $R$ is injective, property (7) is equivalent to property (1).
  If $R$ is univalent, property (8) is equivalent to property (1).
  If $R$ is both injective and univalent, property (9) is equivalent to property (1).
\end{thm}

Furthermore, property (9) of Theorem \ref{theorem.one_strongly_connected} is equivalent to the corresponding equality $R \comp \L \comp R = R \comp R^+$ and property (10) is equivalent to the corresponding equality $R \comp \L \comp R = R^+$.

In \cite{Schmidt2011}, a relation $R$ is called a \emph{simple circuit} if it is univalent, injective and satisfies property (\ref{schmidt-simple-circuit}) of the previous theorem and the condition $R \comp R \rleq \cp{\I}$.
By the previous theorem, this is equivalent to $R$ being a cycle except for the additional condition that $R \comp R$ is irreflexive.
The latter states that no two vertices are mutually connected by an edge; this excludes cycles of length $2$.
The condition also implies that $R$ is irreflexive, which excludes cycles of length $1$, that is, loops.
When undirected graphs are represented as symmetric relations, cycles of length $2$ cannot be distinguished from edges and loops are typically not considered (simple graphs).
For directed graphs, however, we wish to include cycles of length $1$ or $2$ in our reasoning, so we do not require the additional condition.

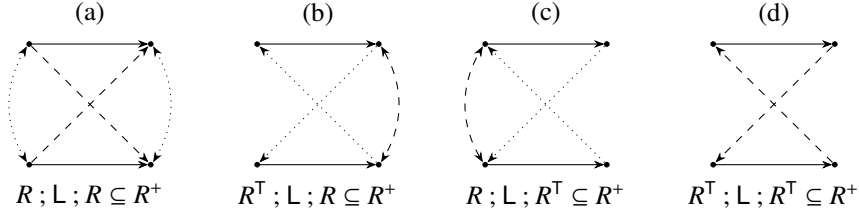
\begin{figure}
  \begin{center}
  \begin{tikzpicture}
    \node at (0.8,2.0) {(a)};
    \draw[fill] (0.0,0.0) circle (0.3mm);
    \draw[fill] (1.6,0.0) circle (0.3mm);
    \draw[fill] (0.0,1.6) circle (0.3mm);
    \draw[fill] (1.6,1.6) circle (0.3mm);
    \path[->] (0.0,0.0) edge (1.57,0.0)
              (0.0,1.6) edge (1.57,1.6);
    \path[<->,dotted] (-0.03,0.03) edge [bend left] (-0.03,1.57)
                      (1.63,1.57) edge [bend left] (1.63,0.03);
    \path[->,dashed] (0.03,0.03) edge (1.57,1.57)
                     (0.03,1.57) edge (1.57,0.03);
    \node at (0.8,-0.4) {$R \comp \L \comp R \rleq R^+$};
    \node at (3.8,2.0) {(b)};
    \draw[fill] (3.0,0.0) circle (0.3mm);
    \draw[fill] (4.6,0.0) circle (0.3mm);
    \draw[fill] (3.0,1.6) circle (0.3mm);
    \draw[fill] (4.6,1.6) circle (0.3mm);
    \path[->] (3.0,0.0) edge (4.57,0.0)
              (3.0,1.6) edge (4.57,1.6);
    \path[<->,dashed] (4.63,1.57) edge [bend left] (4.63,0.03);
    \path[->,dotted] (4.57,0.03) edge (3.03,1.57)
                     (4.57,1.57) edge (3.03,0.03);
    \node at (3.8,-0.4) {$\trans{R} \comp \L \comp R \rleq R^+$};
    \node at (6.8,2.0) {(c)};
    \draw[fill] (6.0,0.0) circle (0.3mm);
    \draw[fill] (7.6,0.0) circle (0.3mm);
    \draw[fill] (6.0,1.6) circle (0.3mm);
    \draw[fill] (7.6,1.6) circle (0.3mm);
    \path[->] (6.0,0.0) edge (7.57,0.0)
              (6.0,1.6) edge (7.57,1.6);
    \path[<->,dashed] (5.97,0.03) edge [bend left] (5.97,1.57);
    \path[->,dotted] (7.57,0.03) edge (6.03,1.57)
                     (7.57,1.57) edge (6.03,0.03);
    \node at (6.8,-0.4) {$R \comp \L \comp \trans{R} \rleq R^+$};
    \node at (9.8,2.0) {(d)};
    \draw[fill] (9.0,0.0) circle (0.3mm);
    \draw[fill] (10.6,0.0) circle (0.3mm);
    \draw[fill] (9.0,1.6) circle (0.3mm);
    \draw[fill] (10.6,1.6) circle (0.3mm);
    \path[->] (9.0,0.0) edge (10.57,0.0)
              (9.0,1.6) edge (10.57,1.6);
    \path[->,dashed] (10.57,0.03) edge (9.03,1.57)
                     (10.57,1.57) edge (9.03,0.03);
    \node at (9.8,-0.4) {$\trans{R} \comp \L \comp \trans{R} \rleq R^+$};
  \end{tikzpicture}
  \end{center}
  \caption{$R$ is strongly connected in (b)--(d), but not strongly connected in (a)}
  \label{figure.start-end}
\end{figure}

Note that properties (3), (7), (8) and (10) of Theorem \ref{theorem.one_strongly_connected} have identical right-hand sides and similar left-hand sides.
Despite this, property (10) is weaker than the others in the following sense, which is illustrated in Figure \ref{figure.start-end}.
It shows, for each of the four properties, a pair of edges of $R$ as solid lines.
We know that these edges exist by considering the left-hand sides of the four properties; the edges may be identical or have the same source vertex or target vertex.
Dashed lines indicate non-empty paths whose existence follows by applying the respective property to these edges.
We know that these paths are not empty because the right-hand side of each property is $R^+$.
From this, the existence of (possibly empty) paths indicated by dotted lines follows if $R$ is univalent (c) or injective (b) or both (a).
The existing paths in (b)--(d) allow us to show that the two edges are in the same strongly connected component of $R$ thereby creating a cycle.
This does not work in (a) even if $R$ is both injective and univalent.
However, with the stronger assumption $R \comp \L \comp R \rleq R \comp R^+$, which is property (9) of the previous theorem, the dashed paths in (a) would have length $2$ or more.
Hence the dotted paths in (a) would be non-empty (and therefore the two edges would be different), so the argument of (b) or (c) could be applied to show that the edges belong to a cycle.

Sometimes we wish to include cycles in reasoning about paths with start or end points.
A path is \emph{backward finite} if it is a cycle or backward terminating [\pfinite\poutinf\pcycle\pempty].
A path is \emph{forward finite} if it is a cycle or forward terminating [\pfinite\pininf\pcycle\pempty].
A path is \emph{finite} if it is a cycle or terminating [\pfinite\pcycle\pempty].
The following result shows how to express each of these properties as an inequality.

\begin{thm}
  Let $R$ be a relation.
  Then the following properties are equivalent and, if $R$ is a path, characterise backward finiteness.
  \begin{enumerate}
  \item $R^* = \trans{R}^*$ or $R \rleq \cp{\L \comp R} \comp R \comp \L$
  \item $R \rleq \cp{\L \comp R} \comp R \comp \L \uni \trans{R}^*$
  \end{enumerate}
  Moreover, the following properties are equivalent and, if $R$ is a path, characterise forward finiteness.
  \begin{enumerate}
  \setcounter{enumi}{2}
  \item $R^* = \trans{R}^*$ or $R \rleq \L \comp R \comp \cp{R \comp \L}$
  \item $R \rleq \L \comp R \comp \cp{R \comp \L} \uni \trans{R}^*$
  \end{enumerate}
  Finally, the following properties are equivalent and, if $R$ is a path, characterise finiteness.
  \begin{enumerate}
  \setcounter{enumi}{4}
  \item $R^* = \trans{R}^*$ or $R \rleq \cp{\L \comp R} \comp R \comp \L \isc \L \comp R \comp \cp{R \comp \L}$
  \item $R \rleq (\cp{\L \comp R} \comp R \comp \L \isc \L \comp R \comp \cp{R \comp \L}) \uni \trans{R}^*$
  \end{enumerate}
\end{thm}

It follows that a path is backward finite if and only if its converse is forward finite.
Moreover a path is finite if and only if its converse is finite.

We conclude this section with a number of useful facts about cycles.
The first result shows that any terminating path can be extended to a cycle by connecting its end point with its start point.

\begin{thm}
\label{path_edge_equals_cycle}
  Let $R$ be a terminating path.
  Then $R \uni \ept{R} \comp \trans{\spt{R}}$ is a cycle.
\end{thm}

Conversely, the next result shows that any non-empty cycle becomes a terminating path if an edge is removed.

\begin{thm}
  Let $R$ be a cycle and let $s, e$ be points with $e \comp \trans{s} \rleq R$.
  Then $R \isc \cp{e \comp \trans{s}}$ is a terminating path with $\spt{R \isc \cp{e \comp \trans{s}}} \rleq s$ and $\ept{R \isc \cp{e \comp \trans{s}}} \rleq e$.
  If $s \neq e$, the last two inequalities can be strengthened to equalities.
\end{thm}

The final result of this section shows how to join two paths with suitable start and end points to a cycle.
It uses the assumption of Theorem \ref{thm:path_concat} to require that the two paths do not overlap.

\begin{thm}
  Let $R, S$ be terminating paths with $\spt{R} = \ept{S}$ and $\spt{S} = \ept{R}$.
  Then $R \uni S$ is a cycle if $R \comp \L \isc (\trans{R} \comp \L \uni S \comp \L) \isc \trans{S} \comp \L = \O$.
\end{thm}

\section{Application: Verifying the Correctness of Graph Algorithms}
\label{section.verification}

Using the relation-algebraic characterisations of different kinds of paths derived in the previous sections, we now verify the correctness of three basic graph algorithms.
Previous work has shown that relation-algebraic reasoning can be applied for program transformation and verification in general and in the context of graph algorithms; for example, see \cite{FriasAguayoNovak1993,BackhouseEtAl94,Berghammer1999,BerghammerStruth2010,BerghammerEtAl2014,Guttmann2018b}.
The aim of this section is to show that our theory of paths integrates well with such arguments.

Our correctness proofs use the well-known assertion-based program verification technique based on Hoare logic \cite{Hoare1969}.
The algorithms are expressed as while-programs with variables whose values are elements of a relation algebra.
Correctness of the algorithms is stated by preconditions and postconditions, which are relation-algebraic formulas.
The first task in proving the correctness of a while-program is to provide for each while-loop an invariant that holds throughout the execution of the loop; invariants are again relation-algebraic formulas.
We then need to prove that each loop invariant
\begin{enumerate}
\item is established from the precondition before the while-loop,
\item is maintained by each iteration of the while-loop, and
\item implies the required postcondition after the while-loop.
\end{enumerate}
This proves partial correctness of the program: if it starts in a state satisfying the precondition and terminates, the postcondition will hold in the final state.

For all our examples, we briefly describe the algorithm itself, and present the pre- and postconditions as well as the loop invariants.
The verification conditions listed above are automatically generated from this information by Isabelle/HOL's Hoare logic library \cite{Nipkow1998,Nipkow2002}.
The proofs of these obligations use relation-algebraic reasoning similar to the proofs of other results in this paper; they can be found in the Isabelle/HOL theory files.

We have also proved termination of the algorithms using a total-correctness Hoare logic library discussed in \cite{Guttmann2018c}.
To this end, each while-loop has to be annotated with a variant or bound function; see the Isabelle/HOL theories for details.
We only discuss partial correctness below; the termination proofs use the additional assumption that graphs have finitely many vertices.

\subsection{Construction of a Path}
\label{ComPath}

Our first example is a Greedy algorithm that constructs a path from a vertex $x$ to a different vertex $y$ of a directed acyclic graph $D$.
We assume that a path between these vertices exists and, moreover, conditions that ensure the Greedy algorithm will find one without searching.
See \cite{BerghammerHoffmann2001,BerghammerHoffmann2001b} for relational implementations of depth-first search and breadth-first search.

We use the predicate $\Ispoint(p)$ to specify that the relation $p$ is a point in the relation-algebraic sense as defined in Section \ref{subsection.points}.

The algorithm maintains a relation $W$, which is the path that is constructed backwards from $y$ towards $x$.
As soon as $W$ forms a path from $x$ to $y$ the algorithm terminates with result $W$ (Line \ref{alg1.line.10} in Algorithm \ref{alg1}).
The algorithm works as follows: first, it chooses a predecessor $p$ of $y$ (Line \ref{alg1.line.2}) and initialises $W$ to be the edge from $p$ to $y$, which is a path (Line \ref{alg1.line.3}).
As long as the start point $q$ of $W$ is different from $x$ (Line \ref{alg1.line.5}), the algorithm chooses a predecessor $p$ of $q$ (Line \ref{alg1.line.6}) and extends the relation $W$ by the edge from $p$ to $q$ (Line \ref{alg1.line.7}).

For the selection of predecessors we use an operation $\Point(v)$ that (deterministically) chooses a point contained in a non-empty vector $v$.
The existence of such a point follows from the point axiom.
To reason about the operation $\Point$, we assume it satisfies the following axioms:
\[
  \Point(v) \rleq v
  \qquad
  v \neq \O \implies \Ispoint(\Point(v))
\]
The inequality states that the chosen relation is contained in $v$, and the implication states that the chosen relation is a point.
The operation $\Point$ is deterministic in the sense that it always produces the same result for a given argument and hence may be modelled as a function in a relation algebra.
There may be several different implementations of a deterministic operation satisfying these axioms; any one works fine since our reasoning only uses the properties stated by or following from the axioms.

Using $\Point$ we can formally describe the construction of a path in Algorithm \ref{alg1}.

\begin{algorithm}
  \caption{Constructing a path}
  \label{alg1}
  \begin{algorithmic}[1]
    \State $\Input ~ D, x, y$                   \label{alg1.line.1}
    \State $p \gets \Point(D \comp y)$          \label{alg1.line.2}
    \State $W \gets p \comp \trans{y}$          \label{alg1.line.3}
    \State $q \gets p$                          \label{alg1.line.4}
    \While{$q \neq x$}                          \label{alg1.line.5}
      \State $p \gets \Point(D \comp q)$        \label{alg1.line.6}
      \State $W \gets W \uni p \comp \trans{q}$ \label{alg1.line.7}
      \State $q \gets p$                        \label{alg1.line.8}
    \EndWhile                                   \label{alg1.line.9}
    \State $\Output ~ W$                        \label{alg1.line.10}
  \end{algorithmic}
\end{algorithm}

We show correctness of Algorithm \ref{alg1} if the input satisfies the following preconditions, whose conjunction is denoted by $\Pre(D,x,y)$:
\[
  D^+ \rleq \cp{\I}
  \qquad\quad
  \Ispoint(x)
  \qquad\quad
  \Ispoint(y)
  \qquad\quad
  x \neq y
  \qquad\quad
  D^* \comp y \rleq \trans{D}^* \comp x
\]
The first inequality states that $D$ is acyclic.
The following three conditions specify that $x$ and $y$ are distinct vertices.
The last inequality uses two vectors: $D^* \comp y$ contains the transitive predecessors of $y$, which are the vertices from which $y$ is reachable, and $\trans{D}^* \comp x$ contains the transitive successors of $x$, which are the vertices reachable from $x$.
The inequality expresses that every transitive predecessor of $y$ is a transitive successor of $x$, which implies that there is a path from $x$ to $y$.
Moreover, it follows that there is a path from $x$ to any transitive predecessor of $y$, which is the reason why the above Greedy algorithm works without searching.

The postcondition $\Post(D,x,y,W)$ of the algorithm is the conjunction of the following relation-algebraic formulas:
\[
  W \rleq D
  \qquad\quad
  \Termpath(W)
  \qquad\quad
  x = \spt{W}
  \qquad\quad
  y = \ept{W}
\]
The first condition guarantees that the output $W$ of Algorithm \ref{alg1} is contained in the graph $D$.
The second condition states that $W$ is a terminating path as defined in Section \ref{section.start-end-points}.
The remaining two conditions ensure that the constructed path starts in $x$ and ends in $y$.

The loop invariant $\Inv(D,x,y,W,q)$ used to prove partial correctness of Algorithm \ref{alg1} is the conjunction of the following formulas:
\begin{align*}
    & D^+ \rleq \cp{\I}
  & & \Ispoint(x)
  & & \Ispoint(y)
  & & \Ispoint(q)
  & & D^* \comp q \rleq \trans{D}^* \comp x
  \\
    & W \rleq D
  & & \Termpath(W)
  & & q = \spt{W}
  & & y = \ept{W}
\end{align*}
The first line shows that the invariant contains most of $\Pre(D,x,q)$.
Maintaining the precondition throughout the while-loop is trivial since it uses only input variables, whose values do not change.
Nevertheless these conditions need to be formally part of the loop invariant since they are not only necessary for establishing the rest of the invariant but also to maintain it.
Moreover, the invariant contains $\Post(D,q,y,W)$ as shown in the second line.

From this invariant, $\Post(D,x,y,W)$ immediately follows using the negated loop condition $q = x$, which holds after the while-loop terminates.

\subsection{Topological Sorting}
\label{ComTopsort}

In our second example we look at topological sorting: given a directed acyclic graph $R$, the problem is to construct a linear order of its vertices that contains $x$ before $y$ for each edge $(x,y)$ of the graph.
If the input graph models dependencies between tasks, the output is a linear schedule of the tasks that respects all dependencies.
We represent the linear order of vertices as a path, in which the sequence of vertices gives the schedule.

A simple algorithm based on \cite{Kahn1962} for finding a topological sort $W$ works as follows.
It first picks a vertex without a predecessor (Line \ref{alg2.line.3} in Algorithm \ref{alg2}), which exists since the input $R$ is acyclic.
This vertex is marked (Line \ref{alg2.line.4}) and used as the start point of the path to be constructed.
As long as there are unmarked vertices (Line \ref{alg2.line.5}), the algorithm picks one that does not have an unmarked predecessor (Line \ref{alg2.line.6}).
As before, such a vertex exists since any subgraph of $R$ is acyclic, too.
The edge from the current end point $q$ of $W$ to the chosen vertex is then added to the path (Line \ref{alg2.line.7}).
The selected vertex is marked (Line \ref{alg2.line.9}).

The relational program given in Algorithm \ref{alg2} implements this procedure.
It represents the set of marked vertices as the vector $v$; a vertex is marked if and only if the corresponding (relational) point is contained in $v$.
Hence $\cp{v}$ contains the unmarked vertices, and the condition $v \neq \L$ in Line \ref{alg2.line.5} checks if there is still an unmarked vertex.

\begin{algorithm}
  \caption{Topological sorting}
  \label{alg2}
  \begin{algorithmic}[1]
    \State $\Input ~ R$                          \label{alg2.line.1}
    \State $W \gets \O$                          \label{alg2.line.2}
    \State $q \gets \Point(\Min(R,\L))$          \label{alg2.line.3}
    \State $v \gets q$                           \label{alg2.line.4}
    \While{$v \neq \L$}                          \label{alg2.line.5}
       \State $p \gets \Point(\Min(R,\cp{v}))$   \label{alg2.line.6}
       \State $W \gets W \uni q \comp \trans{p}$ \label{alg2.line.7}
       \State $q \gets p$                        \label{alg2.line.8}
       \State $v \gets v \uni p$                 \label{alg2.line.9}
    \EndWhile                                    \label{alg2.line.10}
    \State $\Output ~ W$                         \label{alg2.line.11}
  \end{algorithmic}
\end{algorithm}

To find vertices without unmarked predecessors, Algorithm \ref{alg2} uses the operation $\Min(S,w) = w \isc \cp{\trans{S} \comp w}$, where $S$ is a relation and $w$ is a vector \cite{SchmidtStroehlein1993}.
It yields the $S$-minimal elements of the set represented by $w$.
Hence $\Min(R,\L)$ in Line \ref{alg2.line.3} gives the vertices without predecessors in the whole graph, and $\Min(R,\cp{v})$ in Line \ref{alg2.line.6} gives the unmarked vertices without any unmarked predecessors.
As in Algorithm \ref{alg1}, we use the $\Point$ operation to select one of these vertices.

Algorithm \ref{alg2} works if $R$ is acyclic and finite; otherwise one of the $\Point$-operations in Lines \ref{alg2.line.3} and \ref{alg2.line.6} would fail.
To avoid this kind of failure and to ensure that there is no infinite descending chain of unmarked vertices, we require that the input relation $R$ is \emph{well-founded} (also called \emph{regressively finite}).
In relation algebras this is expressed by
\[
  w \rleq \trans{R} \comp w \Rightarrow w = \O
\]
for all vectors $w$ \cite{SchmidtStroehlein1993}.
This property is the only precondition $\Pre(R)$ of the algorithm.

We now consider the postcondition.
The relation $W$ stores the constructed topological sort as a terminating path with a start point and an end point.
To express that $W$ preserves the dependencies in $R$ we use the condition $R \rleq W^+$, that is, there must be a path in $W$ from the source $x$ to the target $y$ of every edge $(x,y) \in R$.
Finally, the path $W$ has to contain all vertices of $R$.
If $R$ has two or more vertices this can be specified as $(W \uni \trans{W}) \comp \L = \L$, where the left-hand side is the vector of all points in $W$.
Alternative expressions for this vector are $\trans{W} \comp \L \uni \spt{W}$ and $W \comp \L \uni \ept{W}$.
If $R$ contains only a single vertex, $W$ is the empty path, so this specification does not work.
We can either add a precondition $\I \neq \L$ stating that $R$ has at least two vertices or modify the condition by replacing $\L$ with $\cp{\I} \comp \L$ to handle this special case.
Using the latter option, we obtain the postcondition $\Post(R,W)$ as the conjunction of the following formulas:
\[
  R \rleq W^+
  \qquad\quad
  \Termpath(W)
  \qquad\quad
  (W \uni \trans{W}) \comp \L = \cp{\I} \comp \L
\]

The loop invariant that allows us to prove partial correctness with respect to the precondition $\Pre(R)$ and the postcondition $\Post(R,W)$ is the conjunction of the following formulas:
\begin{align*}
    & \Pre(R)
  & & R \isc v \comp \trans{v} \rleq W^+
  & & \Termpath(W)
  \\
    & \Ispoint(q)
  & & q \rleq v
  & & W = \O \vee q = \ept{W}
  \\
    & v = v \comp \L
  & & W \comp \L = v \isc \cp{q}
  & & R \comp v \rleq v
\end{align*}
The first line contains the precondition, which is maintained trivially since $R$ never changes, and the first two parts of the postcondition; note that the formula $R \isc v \comp \trans{v} \rleq W^+$ is a generalisation of $R \rleq W^+$ that restricts $R$ to the subgraph induced by the set of marked vertices.
The remaining properties (second and third lines) are auxiliary properties necessary to maintain the first three formulas of the loop invariant and to ensure the postcondition $(W \uni \trans{W}) \comp \L = \cp{\I} \comp \L$.
The second line characterises the relation $q$ as a marked vertex, which is the end point of $W$ except when $W$ is empty at the beginning of Algorithm \ref{alg2}.
The third line characterises the relation $v$ as the set of vertices of $W$ and ensures that all predecessors of these marked vertices are marked, too.

\subsection{Construction of a Non-empty Cycle}
\label{ComCycle}

Our last application is a correctness proof of an algorithm that constructs a non-empty cycle for a given directed graph $R$.
As precondition $\Pre(R)$ we assume that $R$ is not acyclic, that is, it contains at least one cycle:
\[
  R^+ \isc \I \neq \O
\]
This is the only condition needed for input $R$.

Algorithm \ref{alg3} shows the relational program.
It starts by picking two points, using the operation $\Point$ again.
The first point $y$ is selected in Line \ref{alg3.line.2} as a vertex lying on an arbitrary cycle of $R$.
The existence of such a point is guaranteed by $\Pre(R)$.
The second point $x$ is selected in Line \ref{alg3.line.3} as a direct successor of $y$ (since $x \rleq \trans{R} \comp y$) that is also a transitive predecessor of $y$ (since $x \rleq R^* \comp y$).
These conditions ensure that $x$ and $y$ lie on a cycle of $R$.
Lines \ref{alg3.line.4} and \ref{alg3.line.5} cover the case that the chosen points are identical.
This means that the edge $y \comp \trans{x} = x \comp \trans{x}$ is a loop and hence a cycle in $R$, so the algorithm can terminate.

If $x$ and $y$ do not coincide, the algorithm progresses in three steps.
Lines \ref{alg3.line.7}--\ref{alg3.line.13} construct a directed tree $D \rleq R$ with root $x$ in which $y$ is a leaf.
Lines \ref{alg3.line.14}--\ref{alg3.line.21} select a path $W$ in $D$ that connects $x$ with $y$.
This path $W$ together with the edge from $y$ to $x$ gives the required cycle in Line \ref{alg3.line.22}.

In the first step, the directed tree $D$ is constructed as follows.
Line \ref{alg3.line.7} initialises the relation $D$ as the empty tree.
The while-loop maintains the vector $v$ that contains all vertices of the tree.
At the start of the loop, $v$ contains only the point $x$ which is the root of $D$; see Line \ref{alg3.line.8}.
As long as the vertex $y$ has not been reached, which is checked in Line \ref{alg3.line.9}, the algorithm chooses an edge $e$ of $R$ that goes from a vertex in $v$ to a vertex outside $v$ in Line \ref{alg3.line.10}.
The edge is added to the tree in Line \ref{alg3.line.11} and its end point is added to $v$ in Line \ref{alg3.line.12}.

An edge is a relation consisting of a single pair and can therefore be represented by an atom.
We use the predicate $\Isatom(a)$ to specify that the relation $a$ is an atom in the relation-algebraic sense as defined in Section \ref{subsection.points}.
Similarly to points, we use an operation $\Atom(x)$ that (deterministically) chooses an atom contained in a non-empty relation $x$.
To reason about the operation $\Atom$, we assume it satisfies the following axioms:
\[
  \Atom(x) \rleq x
  \qquad
  x \neq \O \implies \Isatom(\Atom(x))
\]
Since the while-loop adds edges leaving the set $v$ and since it is known that $y$ is reachable from $x$, the vertex $y$ will eventually be included in $v$, so $D$ will contain a path from $x$ to $y$.

By construction, $D$ is acyclic and satisfies $D^* \comp y \rleq \trans{D}^* \comp x$.
As a consequence we can use Algorithm \ref{alg1} to determine a path from $x$ to $y$; Lines \ref{alg3.line.14}--\ref{alg3.line.21} of Algorithm \ref{alg3} are identical to Lines \ref{alg1.line.2}--\ref{alg1.line.9} of Algorithm \ref{alg1}.
When the second while-loop of Algorithm \ref{alg3} terminates, $W$ contains a terminating path from $x$ to $y$.
Line \ref{alg3.line.22} adds to this path the edge from $y$ to $x$ to obtain a cycle $C$.
By the choice of $x$ and $y$ in Lines \ref{alg3.line.2} and \ref{alg3.line.3} this edge is contained in $R$.

\begin{algorithm}
  \caption{Constructing a cycle}
  \label{alg3}
  \begin{algorithmic}[1]
    \State $\Input ~ R$                                         \label{alg3.line.1}
    \State $y \gets \Point((R^+ \isc \I) \comp \L)$             \label{alg3.line.2}
    \State $x \gets \Point(R^* \comp y \isc \trans{R} \comp y)$ \label{alg3.line.3}
    \If{$x = y$}                                                \label{alg3.line.4}
      \State $C \gets y \comp \trans{x}$                        \label{alg3.line.5}
    \Else                                                       \label{alg3.line.6}
      \State $D \gets \O$                                       \label{alg3.line.7}
      \State $v \gets x$                                        \label{alg3.line.8}
      \While{$\neg (y \rleq v)$}                                \label{alg3.line.9}
        \State $e \gets \Atom(v \comp \trans{\cp{v}} \isc R)$   \label{alg3.line.10}
        \State $D \gets D \uni e$                               \label{alg3.line.11}
        \State $v \gets v \uni \trans{e} \comp \L$              \label{alg3.line.12}
      \EndWhile                                                 \label{alg3.line.13}
      \State $p \gets \Point(D \comp y)$                        \label{alg3.line.14}
      \State $W \gets p \comp \trans{y}$                        \label{alg3.line.15}
      \State $q \gets p$                                        \label{alg3.line.16}
      \While{$q \neq x$}                                        \label{alg3.line.17}
        \State $p \gets \Point(D \comp q)$                      \label{alg3.line.18}
        \State $W \gets W \uni p \comp \trans{q}$               \label{alg3.line.19}
        \State $q \gets p$                                      \label{alg3.line.20}
      \EndWhile                                                 \label{alg3.line.21}
      \State $C \gets W \uni y \comp \trans{x}$                 \label{alg3.line.22}
    \EndIf                                                      \label{alg3.line.23}
    \State $\Output ~ C$                                        \label{alg3.line.24}
  \end{algorithmic}
\end{algorithm}

We now discuss the postcondition of Algorithm \ref{alg3}.
It constructs a non-empty cycle $C$ that is also a subgraph of $R$.
These conditions immediately translate into the following formulas, whose conjunction is the postcondition $\Post(R,C)$:
\[
  C \neq \O
  \qquad
  \Cycle(C)
  \qquad
  C \rleq R
\]
Here the condition $\Cycle(C)$ specifies that $C$ is a cycle as defined in Section \ref{section.cycles}.

We conclude this section by discussing the invariant used to prove partial correctness of Algorithm \ref{alg3} with respect to $\Pre(R)$ and $\Post(R,C)$.
As before, Isabelle/HOL's Hoare logic tactic splits the correctness proof into a number of verification conditions.
For example, the if-statement is split into two goals, one for each case.
For the case $x = y$ the postcondition is easy to establish since the loop on this vertex is immediately available.

The case $x \neq y$ requires more effort.
To reuse Algorithm \ref{alg1} we have to show that its precondition $\Pre(D,x,y)$ holds, which is the conjunction of the following formulas:
\[
  D^+ \rleq \cp{\I}
  \qquad\quad
  \Ispoint(x)
  \qquad\quad
  \Ispoint(y)
  \qquad\quad
  x \neq y
  \qquad\quad
  D^* \comp y \rleq \trans{D}^* \comp x
\]
To establish these conditions we use the conjunction of the following formulas as the invariant for the first while-loop in Lines \ref{alg3.line.9}--\ref{alg3.line.13}:
\begin{align*}
    & \Ispoint(x)
  & & \Ispoint(y)
  & & x \neq y
  & & y \rleq \trans{R}^* \comp x
  & & y \comp \trans{x} \rleq R
  \\
    & D \rleq R
  & & D^+ \rleq \cp{\I}
  & & D \comp \trans{D} \rleq \I
  & & D \rleq v \comp \trans{v}
  & & v = v \comp \L
  \\
    & x \comp \trans{v} \rleq D^*
  & & D \comp x = \O
  & & v = x \uni \trans{D} \comp \L
\end{align*}
The first line lists simple conditions that follow immediately from the selection of $x$ and $y$ in Lines \ref{alg3.line.2} and \ref{alg3.line.3} of Algorithm \ref{alg3}.
These properties are needed later and have to be maintained throughout the entire proof, which is not difficult since $R$, $x$ and $y$ are not modified in Lines \ref{alg3.line.4}--\ref{alg3.line.23}.

The first three inequalities of the second line state that $D$ is an injective acyclic subgraph of $R$, that is, a forest in $R$.
The remaining two conditions in this line state that $v$ is a vector and that $D$ contains only edges between vertices in the set represented by $v$.

The first two conditions in the third line specify that $x$ is the root of $D$, that is, that all vertices in $v$ are reachable from $x$ in $D$ and that $x$ has no predecessors.
This implies that $D$ is a tree with root $x$.
Finally, $v = x \uni \trans{D} \comp \L$ states that $v$ contains $x$ and the target vertices of all edges in $D$.

These invariants imply that $D^* \comp y \rleq \trans{D}^* \comp x$ holds after the first while-loop, so we can use Algorithm \ref{alg1} afterwards.
By the correctness of Algorithm \ref{alg1} we know that its postcondition $\Post(D,x,y,W)$ holds after the second while-loop of Algorithm \ref{alg3}.
This postcondition is:
\[
  W \rleq D
  \qquad\quad
  \Termpath(W)
  \qquad\quad
  x = \spt{W}
  \qquad\quad
  y = \ept{W}
\]
Theorem \ref{path_edge_equals_cycle} then implies that $C$ as constructed in Line \ref{alg3.line.22} is a cycle.
The remaining two postconditions of Algorithm \ref{alg3}, that is, $C \neq \O$ and $C \rleq R$ are easy to prove using $x \neq y$ and $W \rleq D$ and $D \rleq R$.

\section{Paths with Roots}
\label{section.roots}

In the previous sections a path was described by a relation only; in Section \ref{section.start-end-points} we have discussed conditions under which paths have a start point and/or an end point.
In this section we describe paths together with a designated root.
A root is a vertex of the graph represented by a (relational) point.
Our main results are equivalences between the definitions of paths with and without roots.
Most of these equivalences only hold if the point axiom is assumed.

We represent a path with root by two relations $R$ and $p$ such that $R$ is injective and univalent and $p$ is a point.
The relations $R$ and $p$ will satisfy further conditions depending on the kind of path that is represented.

\subsection{Paths}

We start with a result that characterises all paths [\pfinite\pininf\poutinf\pdblinf\pcycle\pempty].
The condition $p \comp R \rleq R^* \uni \trans{R}^*$ states that any end vertex of an edge in $R$ must be reachable from the vertex $p$ by going forward in $R$ or by going backward in $R$.

\begin{thm}
  Let $R$ be an injective and univalent relation.
  Then $R$ is a path if and only if there exists a point $p$ such that $p \comp R \rleq R^* \uni \trans{R}^*$.
\end{thm}

If the path is not empty, we can guarantee that the vertex $p$ of the previous theorem is contained in the path.
This is stated by the additional condition $p \rleq (R \uni \trans{R}) \comp \L$ of the following characterisation [\pfinite\pininf\poutinf\pdblinf\pcycle].

\begin{cor}
  Let $R$ be an injective and univalent relation.
  Then $R$ is a non-empty path if and only if there exists a point $p$ such that $p \comp R \rleq R^* \uni \trans{R}^*$ and $p \rleq (R \uni \trans{R}) \comp \L$.
\end{cor}

The following theorem shows that a path $R$ is acyclic if $p$ has no predecessors in $R$.

\begin{thm}
  \label{theorem.path_root_acyclic}
  Let $R$ be an injective and univalent relation and let $p$ be a point such that $p \comp R \rleq R^* \uni \trans{R}^*$ and $R \comp p = \O$.
  Then $R$ is acyclic.
\end{thm}

\subsection{Backward Finite Paths}

The following result characterises backward finite paths [\pfinite\poutinf\pcycle\pempty].
The condition $p \comp R \rleq R^+$ states that any end vertex of an edge in $R$ must be reachable from the vertex $p$ by a non-empty path in $R$.
It implies the previous condition $p \comp R \rleq R^* \uni \trans{R}^*$ since $R^+ \rleq R^* \rleq R^* \uni \trans{R}^*$.

\begin{thm}
  Let $R$ be an injective and univalent relation.
  Then $R$ is a backward finite path if and only if there exists a point $p$ such that $p \comp R \rleq R^+$.
\end{thm}

For a point $p$, the condition $p \comp R \rleq R^+$ is equivalent to each of $\trans{R} \comp \L \rleq \trans{R}^+ \comp p$ and $\trans{R} \comp \L = \trans{R}^+ \comp p$.

Again, if the path is not empty, we obtain that it contains $p$, which can now be stated as $p \rleq R \comp \L$ in the following characterisation [\pfinite\poutinf\pcycle].

\begin{cor}
  Let $R$ be an injective and univalent relation.
  Then $R$ is a non-empty backward finite path if and only if there exists a point $p$ such that $p \comp R \rleq R^+$ and $p \rleq R \comp \L$.
\end{cor}

For backward finite paths, Theorem \ref{theorem.path_root_acyclic} can be extended to an equivalence as the following result shows.

\begin{thm}
  Let $R$ be an injective and univalent relation and let $p$ be a point such that $p \comp R \rleq R^+$.
  Then $R \comp p = \O$ if and only if $R$ is acyclic.
\end{thm}

\subsection{Cycles}

The following result characterises non-empty cycles [\pcycle].
The additional condition $p \rleq \trans{R} \comp \L$ states that $p$ is the end vertex of an edge in $R$.
For a point $p$ it is equivalent to $R \comp p \neq \O$.

\begin{thm}
  Let $R$ be an injective and univalent relation.
  Then $R$ is a non-empty cycle if and only if there exists a point $p$ such that $p \comp R \rleq R^+$ and $p \rleq \trans{R} \comp \L$.
\end{thm}

In this case any point $q$ on the cycle -- that is, satisfying $q \rleq R^* \comp p$ -- can take the place of $p$ in the previous theorem -- that is, satisfies $q \comp R \rleq R^+$ and $q \rleq \trans{R} \comp \L$.
Moreover, $p \rleq R^* \comp q$ and $q \rleq R^+ \comp q = R^* \comp q = R^* \comp p = R \comp \L$ follow.
Finally, also $p \comp \trans{R} \rleq \trans{R}^+$ and $p \rleq R \comp \L$ hold, so the previous theorem dualises to the converse cycle.
The condition $q \rleq R^* \comp p$ to make all of this happen for non-empty cycles can equivalently be stated as each of $p \rleq R^* \comp q$ or $R \comp q \neq \O$ or $\trans{R} \comp q \neq \O$ or $q \rleq R \comp \L$ or $q \rleq \trans{R} \comp \L$.

Moreover, it follows that if $R$ and $S$ are non-empty cycles such that $R \rleq S$ then $R = S$.
Another consequence is that the two conditions of the previous theorem can be combined into one inequality if also the empty cycle is allowed [\pcycle\pempty].

\begin{cor}
  Let $R$ be an injective and univalent relation.
  Then $R$ is a cycle if and only if there exists a point $p$ such that $p \comp R \rleq R^+ \isc \trans{R} \comp \L$.
\end{cor}

\subsection{Backward Terminating Paths}

The following result characterises backward terminating paths [\pfinite\poutinf\pempty].
The additional condition $R \comp p = \O$ states that vertex $p$ has no predecessors in $R$.

\begin{thm}
  Let $R$ be an injective and univalent relation.
  Then $R$ is a backward terminating path if and only if there exists a point $p$ such that $p \comp R \rleq R^+$ and $R \comp p = \O$.
\end{thm}

If the relation is not empty, it follows that the root $p$ of the previous theorem is the start point of $R$ [\pfinite\poutinf].

\begin{cor}
  Let $R$ be an injective and univalent relation.
  Then $R$ is a non-empty backward terminating path if and only if $\spt{R}$ is a point such that $\spt{R} \comp R \rleq R^+$.
\end{cor}

\subsection{Terminating Paths}

For characterising terminating paths we additionally need an end point [\pfinite\pempty].
The condition $p \rleq R^* \comp q$ states that vertex $q$ is reachable from vertex $p$ in $R$.
The condition $\trans{R} \comp q = \O$ states that vertex $q$ has no successors in $R$.

\begin{thm}
  Let $R$ be an injective and univalent relation.
  Then $R$ is a terminating path if and only if there exist points $p, q$ such that $p \comp R \rleq R^+$ and $p \rleq R^* \comp q$ and $\trans{R} \comp q = \O$.
\end{thm}

In this case it follows that $R \comp p = \O$.
Moreover, also $q \rleq \trans{R}^* \comp p$ and $q \comp \trans{R} \rleq \trans{R}^+$ follow, so the previous theorem dualises by swapping $p$ and $q$ and taking the converse of $R$.
Finally, $R = \O$ if and only if $p = q$.

If the relation is not empty, the point $q$ of the previous theorem is the end point of $R$ [\pfinite].

\begin{cor}
  Let $R$ be an injective and univalent relation.
  Then $R$ is a non-empty terminating path if and only if $\spt{R}$ and $\ept{R}$ are points such that $\spt{R} \comp R \rleq R^+$ and $\spt{R} \rleq R^* \comp \ept{R}$.
\end{cor}

\section{Conclusion}

We have shown how relation algebras can be used to compactly specify and reason about different kinds of paths in graphs.
We have applied the developed formalism to verify the correctness of simple graph algorithms. 
Proofs of results use equational reasoning instead of point-wise arguments with quantified variables.
This style of reasoning strongly benefits from support by automated and interactive theorem provers.

In this paper we have developed a fundamental theory of paths.
There are several directions of extension.
First, relation algebras can be extended to deal with cardinalities of relations \cite{Kawahara2006,BerghammerHoefnerStucke16,BerghammerEtAl16}; for a path this would simply be the number of edges it contains.
This facilitates, for example, proofs about the complexity of graph algorithms.
Second, different combinations of the defining properties of paths and other properties yield different classes of graphs.
For example, an injective acyclic relation represents a forest; adding connectivity yields trees.
We expect that a number of results are common to the classes and can be derived from fewer properties making them more widely applicable.
Third, relation algebras can be generalised to structures that can represent weighted graphs \cite{Guttmann2018b}.
Using the results of this paper in such a setting would allow us to reason, for example, about shortest-path algorithms.


\begin{thebibliography}{10}

\bibitem{Relation_Algebra-AFP}
A.~Armstrong, S.~Foster, G.~Struth, and T.~Weber.
\newblock Relation algebra.
\newblock {\em Archive of Formal Proofs}, January 2014.
\newblock \url{http://isa-afp.org/entries/Relation_Algebra.html}, Formal proof
  development.

\bibitem{BackhouseEtAl94}
R.~Backhouse, J.~P. H.~W. van~den Eijnde, and A.~J.~M. van Gasteren.
\newblock Calculating path algorithms.
\newblock {\em Science of Computer Programming}, 22(1--2):3--19, 1994.

\bibitem{Berge2001}
C.~Berge.
\newblock {\em The Theory of Graphs}.
\newblock Dover Publications, 2001.

\bibitem{Berghammer1999}
R.~Berghammer.
\newblock Combining relational calculus and the {Dijkstra--Gries} method for
  deriving relational programs.
\newblock {\em Information Sciences}, 119(3--4):155--171, 1999.

\bibitem{BerghammerEtAl16}
R.~Berghammer, N.~Danilenko, P.~H{\"o}fner, and I.~Stucke.
\newblock Cardinality of relations with applications.
\newblock {\em Discrete Mathematics}, 339(12):3089--3115, 2016.

\bibitem{BerghammerHoffmann2001}
R.~Berghammer and T.~Hoffmann.
\newblock Calculating a relational program for transitive reductions of
  strongly connected graphs.
\newblock In H.~de~Swart, editor, {\em Relational Methods in Computer Science
  (RelMiCS 2001)}, volume 2561 of {\em Lecture Notes in Computer Science},
  pages 258--275. Springer, 2001.

\bibitem{BerghammerHoffmann2001b}
R.~Berghammer and T.~Hoffmann.
\newblock Relational depth-first search with applications.
\newblock {\em Information Sciences}, 139(3--4):167--186, 2001.

\bibitem{BerghammerEtAl2014}
R.~Berghammer, P.~H{\"o}fner, and I.~Stucke.
\newblock Automated verification of relational while-programs.
\newblock In P.~H{\"o}fner, P.~Jipsen, W.~Kahl, and M.~E. M{\"u}ller, editors,
  {\em Relational and Algebraic Methods in Computer Science (RAMiCS 2014)},
  volume 8428 of {\em Lecture Notes in Computer Science}, pages 173--190.
  Springer, 2014.

\bibitem{BerghammerHoefnerStucke2015}
R.~Berghammer, P.~H{\"o}fner, and I.~Stucke.
\newblock Tool-based verification of a relational vertex coloring program.
\newblock In W.~Kahl, J.~N. Oliveira, and M.~Winter, editors, {\em Relational
  and Algebraic Methods in Computer Science (RAMiCS 2015)}, volume 9348 of {\em
  Lecture Notes in Computer Science}, pages 275--292. Springer, 2015.

\bibitem{BerghammerHoefnerStucke16}
R.~Berghammer, P.~H{\"o}fner, and I.~Stucke.
\newblock Cardinality of relations and relational approximation algorithms.
\newblock {\em Journal of Logical and Algebraic Methods in Programming},
  85:269--286, 2016.

\bibitem{BerghammerStruth2010}
R.~Berghammer and G.~Struth.
\newblock On automated program construction and verification.
\newblock In C.~Bolduc, J.~Desharnais, and B.~Ktari, editors, {\em Mathematics
  of Program Construction (MPC 2010)}, volume 6120 of {\em Lecture Notes in
  Computer Science}, pages 22--41. Springer, 2010.

\bibitem{Bertot2004}
Y.~Bertot and P.~Cast{\'e}ran.
\newblock {\em Interactive theorem proving and program development}.
\newblock Texts in Theoretical Computer Science. Springer, 2004.

\bibitem{BlanchetteBoehmePaulson2011}
J.~C. Blanchette, S.~B{\"o}hme, and L.~C. Paulson.
\newblock Extending {Sledgehammer} with {SMT} solvers.
\newblock In N.~Bj{\o{}}rner and V.~Sofronie-Stokkermans, editors, {\em
  Automated Deduction: CADE-23}, volume 6803 of {\em Lecture Notes in Computer
  Science}, pages 116--130. Springer, 2011.

\bibitem{Conway1971}
J.~H. Conway.
\newblock {\em Regular Algebra and Finite Machines}.
\newblock Chapman \& Hall, 1971.

\bibitem{CormenLeisersonRivest1990}
T.~H. Cormen, C.~E. Leiserson, and R.~L. Rivest.
\newblock {\em Introduction to Algorithms}.
\newblock MIT Press, 1990.

\bibitem{DangHoefner2008}
H.-H. Dang and P.~H{\"o}fner.
\newblock First-order theorem prover evaluation w.r.t. relation- and {Kleene}
  algebra.
\newblock In R.~Berghammer, B.~M{\"o}ller, and G.~Struth, editors, {\em
  Relations and Kleene Algebra in Computer Science: PhD Programme at
  RelMiCS10/AKA5}, Report 2008-04, pages 48--52. Institut f{\"u}r Informatik,
  Universit{\"a}t Augsburg, 2008.

\bibitem{DesharnaisMoellerStruth2011}
J.~Desharnais, B.~M{\"o}ller, and G.~Struth.
\newblock Algebraic notions of termination.
\newblock {\em Logical Methods in Computer Science}, 7(1:1):1--29, 2011.

\bibitem{Diestel2005}
R.~Diestel.
\newblock {\em Graph Theory}.
\newblock Springer, third edition, 2005.

\bibitem{FosterStruthWeber2011}
S.~Foster, G.~Struth, and T.~Weber.
\newblock Automated engineering of relational and algebraic methods in
  {Isabelle/HOL}.
\newblock In H.~de~Swart, editor, {\em Relational and Algebraic Methods in
  Computer Science}, volume 6663 of {\em Lecture Notes in Computer Science},
  pages 52--67. Springer, 2011.

\bibitem{FriasAguayoNovak1993}
M.~F. Frias, N.~Aguayo, and B.~Novak.
\newblock Development of graph algorithms with fork algebras.
\newblock In {\em XIX Conferencia Latinoamericana de Inform{\'a}tica}, pages
  529--554, 1993.

\bibitem{Glueck2017}
R.~Gl\"uck.
\newblock Algebraic investigation of connected components.
\newblock In P.~H{\"o}fner, D.~Pous, and G.~Struth, editors, {\em Relational
  and Algebraic Methods in Computer Science (RAMiCS 2017)}, volume 10226 of
  {\em Lecture Notes in Computer Science}, pages 109--126. Springer, 2017.

\bibitem{Guttmann2016a}
W.~Guttmann.
\newblock An algebraic approach to computations with progress.
\newblock {\em Journal of Logical and Algebraic Methods in Programming},
  85(4):520--539, 2016.

\bibitem{Guttmann2018b}
W.~Guttmann.
\newblock An algebraic framework for minimum spanning tree problems.
\newblock {\em Theoretical Computer Science}, 744:37--55, 2018.

\bibitem{Guttmann2018c}
W.~Guttmann.
\newblock Verifying minimum spanning tree algorithms with {Stone} relation
  algebras.
\newblock {\em Journal of Logical and Algebraic Methods in Programming},
  101:132--150, 2018.

\bibitem{Harary1969}
F.~Harary.
\newblock {\em Graph Theory}.
\newblock Addison-Wesley Publishing Company, 1969.

\bibitem{HirschHodkinson2002}
R.~Hirsch and I.~Hodkinson.
\newblock {\em Relation Algebras by Games}.
\newblock Elsevier Science B.V., 2002.

\bibitem{Hoare1969}
C.~A.~R. Hoare.
\newblock An axiomatic basis for computer programming.
\newblock {\em Communications of the ACM}, 12(10):576--580/583, 1969.

\bibitem{HoefnerStruth2008}
P.~H{\"o}fner and G.~Struth.
\newblock On automating the calculus of relations.
\newblock In A.~Armando, P.~Baumgartner, and G.~Dowek, editors, {\em Automated
  Reasoning ({IJCAR} 2008)}, volume 5195 of {\em Lecture Notes in Artificial
  Intelligence}, pages 50--66. Springer, 2008.

\bibitem{Huntington1933b}
E.~V. Huntington.
\newblock Boolean algebra. {A} correction.
\newblock {\em Transactions of the AMS}, 35:557--558, 1933.

\bibitem{Huntington1933a}
E.~V. Huntington.
\newblock New sets of independent postulates for the algebra of logic.
\newblock {\em Transactions of the AMS}, 35:274--304, 1933.

\bibitem{Kahn1962}
A.~B. Kahn.
\newblock Topological sorting of large networks.
\newblock {\em Communications of the ACM}, 5(11):558--562, 1962.

\bibitem{Kawahara2006}
Y.~Kawahara.
\newblock On the cardinality of relations.
\newblock In R.~A. Schmidt, editor, {\em Relations and Kleene Algebra in
  Computer Science}, volume 4136 of {\em Lecture Notes in Computer Science},
  pages 251--265. Springer, 2006.

\bibitem{Kozen1994}
D.~Kozen.
\newblock A completeness theorem for {Kleene} algebras and the algebra of
  regular events.
\newblock {\em Information and Computation}, 110(2):366--390, 1994.

\bibitem{MacCaullOrlowska2002}
W.~MacCaull and E.~Or{\l}owska.
\newblock Correspondence results for relational proof systems with application
  to the {Lambek} calculus.
\newblock {\em Studia Logica}, 71(3):389--414, 2002.

\bibitem{Maddux1983}
R.~Maddux.
\newblock A sequent calculus for relation algebras.
\newblock {\em Annals of Pure and Applied Logic}, 25:73--101, 1983.

\bibitem{Maddux2006}
R.~D. Maddux.
\newblock {\em Relation Algebras}.
\newblock Elsevier, 2006.

\bibitem{Prover9}
W.~W. McCune.
\newblock Prover9 and {Mace4}.
\newblock \url{http://www.cs.unm.edu/~mccune/prover9/}.
\newblock (accessed 15/03/2017).

\bibitem{Meng2008}
J.~Meng and L.~C. Paulson.
\newblock Translating higher-order clauses to first-order clauses.
\newblock {\em Journal of Automated Reasoning}, 40(1):35--60, 2008.

\bibitem{Moeller13}
B.~M{\"o}ller.
\newblock Modal knowledge and game semirings.
\newblock {\em The Computer Journal}, 56(1):53--69, 2013.

\bibitem{MoellerEtAl12}
B.~M{\"o}ller, P.~Roocks, and M.~Endres.
\newblock An algebraic calculus of database preferences.
\newblock In J.~Gibbons and P.~Nogueira, editors, {\em Mathematics of Program
  Construction}, volume 7342 of {\em Lecture Notes in Computer Science}, pages
  241--262. Springer, 2012.

\bibitem{Mueller12}
M.~E. M{\"u}ller.
\newblock {\em Relational Knowledge Discovery}.
\newblock Cambridge University Press, 2012.

\bibitem{Ng1984}
K.~C. Ng.
\newblock {\em Relation Algebras with Transitive Closure}.
\newblock PhD thesis, University of California, Berkeley, 1984.

\bibitem{Nipkow1998}
T.~Nipkow.
\newblock Winskel is (almost) right: Towards a mechanized semantics textbook.
\newblock {\em Formal Aspects of Computing}, 10(2):171--186, 1998.

\bibitem{Nipkow2002}
T.~Nipkow.
\newblock Hoare logics in {Isabelle/HOL}.
\newblock In H.~Schwichtenberg and R.~Stein\-br{\"u}ggen, editors, {\em Proof
  and System-Reliability}, pages 341--367. Kluwer Academic Publishers, 2002.

\bibitem{NipkowPaulsonWenzel2002}
T.~Nipkow, L.~C. Paulson, and M.~Wenzel.
\newblock {\em {Isabelle}/{HOL}: A Proof Assistant for {Higher-Order Logic}},
  volume 2283 of {\em Lecture Notes in Computer Science}.
\newblock Springer, 2002.

\bibitem{OkumaKawahara2000}
H.~Okuma and Y.~Kawahara.
\newblock Relational aspects of relational database dependencies.
\newblock {\em Bulletin of Informatics and Cybernetics}, 32(2):91--104, 2000.

\bibitem{PaulsonBlanchette2010}
L.~C. Paulson and J.~C. Blanchette.
\newblock Three years of experience with {Sledgehammer}, a practical link
  between automatic and interactive theorem provers.
\newblock In G.~Sutcliffe, S.~Schulz, and E.~Ternovska, editors, {\em
  International Workshop on the Implementation of Logics ({IWIL} 2010)},
  volume~2 of {\em EPiC Series in Computing}, pages 1--11. EasyChair, 2010.

\bibitem{Pous2016}
D.~Pous.
\newblock {\em Automata for relation algebra and formal proofs}.
\newblock Habilitation {\`a} diriger des recherches, ENS Lyon, 2016.

\bibitem{Schmidt2011}
G.~Schmidt.
\newblock {\em Relational Mathematics}.
\newblock Cambridge University Press, 2011.

\bibitem{Schmidt12}
G.~Schmidt.
\newblock Relational concepts in social choice.
\newblock In W.~Kahl and T.~Griffin, editors, {\em Relational and Algebraic
  Methods in Computer Science}, volume 7560 of {\em Lecture Notes in Computer
  Science}, pages 278--293. Springer, 2012.

\bibitem{Schmidt08}
G.~Schmidt and R.~Berghammer.
\newblock Relational measures and integration in preference modeling.
\newblock {\em Journal of Logic and Algebraic Programming}, 76(1):112--129,
  2008.

\bibitem{SchmidtStroehlein1993}
G.~Schmidt and T.~Str{\"o}hlein.
\newblock {\em Relations and Graphs: Discrete Mathematics for Computer
  Scientists}.
\newblock Springer, 1993.

\bibitem{Schulz2013}
S.~Schulz.
\newblock {System Description: E~1.8}.
\newblock In K.~McMillan, A.~Middeldorp, and A.~Voronkov, editors, {\em Logic
  for Programming Artificial Intelligence and Reasoning (LPAR 19)}, volume 8312
  of {\em Lecture Notes in Computer Science}. Springer, 2013.

\bibitem{ScolloFrancoManca06}
G.~Scollo, G.~Franco, and V.~Manca.
\newblock A relational view of recurrence and attractors in state transition
  dynamics.
\newblock In R.~A. Schmidt, editor, {\em Relations and Kleene Algebra in
  Computer Science}, volume 4136 of {\em Lecture Notes in Computer Science},
  pages 358--372. Springer, 2006.

\bibitem{COST1}
H.~de Swart, E.~Or{\l}owska, G.~Schmidt, and M.~Roubens, editors.
\newblock {\em Theory and Applications of Relational Structures as Knowledge
  Instruments}, volume 2929 of {\em Lecture Notes in Computer Science}.
  Springer, 2003.

\bibitem{COST2}
H.~de Swart, E.~Or{\l}owska, G.~Schmidt, and M.~Roubens, editors.
\newblock {\em {Theory and Applications of Relational Structures as Knowledge
  Instruments II}}, volume 4342 of {\em Lecture Notes in Computer Science}.
  Springer, 2006.

\bibitem{Tarski1941}
A.~Tarski.
\newblock On the calculus of relations.
\newblock {\em The Journal of Symbolic Logic}, 6(3):73--89, 1941.

\bibitem{Tinhofer1976}
G.~Tinhofer.
\newblock {\em Methoden der angewandten Graphentheorie}.
\newblock Springer, 1976.

\bibitem{Wenzel2002}
M.~Wenzel.
\newblock {\em {Isabelle, Isar} -- A Versatile Environment for Human Readable
  Formal Proof Documents}.
\newblock PhD thesis, Technical University Munich, Germany, 2002.

\bibitem{Wright2004}
J.~von Wright.
\newblock Towards a refinement algebra.
\newblock {\em Science of Computer Programming}, 51(1--2):23--45, 2004.

\end{thebibliography}

\end{document}